\documentclass[%
longbibliography,
preprint,
amsmath,amssymb,
aps,
showkeys
]{revtex4-1}
\usepackage{makecell}
\usepackage{multirow}
\usepackage{xcolor}
\usepackage{textcomp}
\usepackage{graphicx}
\usepackage{dcolumn}
\usepackage{bm}

\begin{document}
	
	
	\title{Non-local reconfigurable sparse metasurface: Efficient near-field and far-field wavefront manipulations}
    
    \author{Vladislav Popov$^1$}
	\email{uladzislau.papou@centralesupelec.fr}
	\author{Badreddine Ratni$^2$}
	\author{Shah Nawaz Burokur$^2$}
	\email{sburokur@parisnanterre.fr}
	\author{Fabrice Boust$^{1,3}$}
	\email{fabrice.boust@onera.fr}
	\affiliation{%
	$^1$SONDRA, CentraleSup\'elec, Universit\'e Paris-Saclay,
		F-91190, Gif-sur-Yvette, France
	}%
	\affiliation{%
	$^2$LEME, UPL, Univ Paris Nanterre, F92410 Ville d'Avray, France
	}
	\affiliation{%
$^{3}$DEMR, ONERA, Universit\'e Paris-Saclay, F-91123, Palaiseau, France
	}

	\begin{abstract}
	\noindent
	In recent years, metasurfaces have shown extremely powerful abilities for manipulation of electromagnetic waves.
	However, the local electromagnetic response of conventional metasurfaces yields to an intrinsic performance limitation in terms of efficiency, minimizing their implementation in real-life applications.
	The efficiency of reconfigurable metasurfaces further decreases because of the high density of meta-atoms, reaching 74  meta-atoms per $\lambda^2$ area, incorporating lossy tunable elements.
	To address these problems, we implement strong electromagnetic non-local features in a \textit{sparse} metasurface composed of electronically reconfigurable meta-atoms.
	As a proof-of-concept demonstration, a dynamic sparse metasurface having as few as $8$ meta-atoms per $\lambda^2$ area is experimentally realized in the microwave domain to control wavefronts in both near-field and far-field regions for focusing and beam-forming, respectively. 
	The proposed  metasurface with its sparsity not only facilitates design and fabrication, but also opens the door to high-efficiency real-time reprogrammable functionalities in beam manipulations, wireless power transfer and imaging holography.\\
    \end{abstract}


	\maketitle

\section{Introduction}
From long ago, composite structures have been serving to go beyond characteristics of natural materials. 
A leap further from classical composites was made when the concept of metamaterials emerged as a way to knowingly engineer properties of structures and adapt them to real-life applications.	
In recent years, thin two-dimensional structured surfaces, known as metasurfaces,  have demonstrated unprecedented capabilities in electromagnetic wavefronts tailoring.
Nowadays, there is an increasing research interest in reconfigurable (or tunable) metasurfaces~\cite{Sievenpiper2003,Hum2012_transmitarray,Clemente2013_coding_transmitarray,zhu2014dynamic,cui2014coding,della2014digital,kaina2014shaping,Zheludev2015,Hougne2015,xu2016dynamical,yang2016programmable,li2016transmission,xu2016tunable,Hougne2016,zhou2016focusing,chen2017reconfigurable,li2017electromagnetic,Ratni18,Hougne2018dynamic,komar2018dynamic,Hougne2018,li2019intelligent,ma2019smart,Tretyakov_IntMS2019,Hougne2019,Altug2020,Ratni2020}
due to ongoing fundamental research on time-modulated and nonreciprocal systems~\cite{Alu2015_time_modulated,Alu2018_time_modulated,Zhang2018_time_modulated,Shadrivov2018_time_modulated,Salary2018_Nonreciprocity_time_modulated,Tretyakov2018_Nonreciprocity,Gomez-Diaz2019_Nonreciprocity_time_modulated,Gomez-Diaz2019_Nonreciprocity,Asadchy2019_time_modulated,Ptitcyn2019_time_modulated} on the one hand 
and novel emerging applications on the other hand.
As such, concepts of analogue computing, computer vision,  Internet of Things (IoT), smart homes and smart cities~\cite{Hougne2018,Hougne2018dynamic,li2019intelligent,ma2019smart,Tretyakov_IntMS2019} create an ever increasing demand for compact, versatile and efficient microwave devices to manipulate electromagnetic wavefronts.

Conceptually, the operational principle of conventional metasurfaces is based on adjusting the local phase response of constituting meta-atoms. 
Generally, single-layered metasurfaces do not allow to achieve the required $2\pi$ phase shift for wavefront control in transmission. In most cases, several transmissive layers or a reflective-type metasurface where meta-atoms are patterned on a metal-backed dielectric substrate is exploited.
In a reconfigurable metasurface, the dynamic phase range can be narrowed due to ohmic losses inherent to tunable electronic elements~\cite{Ratni18}.
At microwave frequencies, PIN diodes and varactor diodes are usually implemented in meta-atoms for the reconfigurability mechanism. 
PIN diodes offering two different states (on and off) have been widely used in digital meta-atoms to achieve binary phase states~\cite{Clemente2013_coding_transmitarray,cui2014coding,kaina2014shaping,Hougne2015,xu2016dynamical,yang2016programmable,li2016transmission,Hougne2016,li2017electromagnetic,Hougne2018dynamic,Hougne2018,li2019intelligent,ma2019smart,Hougne2019}. 
Varactor diodes on their side provide a more flexible solution in achieving real-time reconfigurablity since the capacitance value can be modified continuously with a change in applied bias voltage. 
As such, dispersion compensation and dynamical functionality switching has been demonstrated in varactor-based metasurfaces~\cite{Hum2012_transmitarray,zhu2014dynamic,xu2016tunable,chen2017reconfigurable,Ratni18,Tretyakov_IntMS2019,Ratni2020}.
Complicated designs of reconfigurable meta-atoms and ohmic losses of electronic elements make it very challenging, especially in optical and visible domains~\cite{wang2016optically,huang2016gate,komar2018dynamic,Cui2019_Review_optics,he2019tunable}, to bring the reconfigurability mechanism to more advanced concepts of Huygens'~\cite{Grbic2013,Epstein2014_ieee,epstein2016cavity} and bianisotropic metasurfaces~\cite{Asadchy2016,Epstein2016_ieee,Epstein2016_prl}.

While it can be necessary to consider complex designs of reconfigurable meta-atoms for full wavefront control, there are fundamental limitations on the performance of conventionnal metasurfaces as a result of the local normal power flow conservation condition ~\cite{Epstein2014_ieee,Epstein2016_ieee,Asadchy2016,Alu2016}.
Indeed, metasurfaces are theoretically considered as \textit{continuous} surface impedances~\cite{Holloway2003_GSTC,Grbic2013,Epstein:16} and, being of deeply subwavelength thickness, they effectively realize an abrupt discontinuity of the electromagnetic field while forcing the equality between entering and leaving \textit{local} power flows~\cite{Epstein2014_ieee,Epstein2016_ieee}.
In the case of an impenetrable (reflective) metasurface, only the tangential component of power flow exists in its proximity.
A theoretical solution proposed to overcome this limitation
suggests implementing a \textit{strongly non-local} metasurface~\cite{Epstein2016_prl,Tretyakov2017_NLM,Kwon2018_NLM}, which implies existence of a substantial interaction between distant parts of the metasurface via surface waves propagating along it.
So far, strong non-locality has never been implemented in conventional reconfigurable metasurfaces, making therefore their efficiency suffer from dissipation of energy in parasitic directions.
The efficiency further decreases because of the high density of meta-atoms, reaching $74$ meta-atoms per $\lambda^2$ area~\cite{Sievenpiper2003}, incorporating tunable elements  that bring additional ohmic losses to the system. 
Taking the above statements in consideration, the contribution of this study is twofold. 
First, we design and experimentally validate a reconfigurable \textit{sparse} metasurface requiring less tunable elements compared to their traditional counterparts and possessing intrinsic strongly  non-local response~\cite{Alu2017_metagr,Epstein2017_mtg,Popov2019}.
Secondly, direct evidences of strongly non-local behavior of a reconfigurable metasurface are demonstrated for the first time. 
The developed approach represents a radically new paradigm to dynamically manipulate electromagnetic wavefronts and allows us to experimentally demonstrate extreme examples of far-field wave manipulation and subdiffraction near-field focusing.

\begin{figure*}[tb]
	\includegraphics[width=0.99\linewidth]{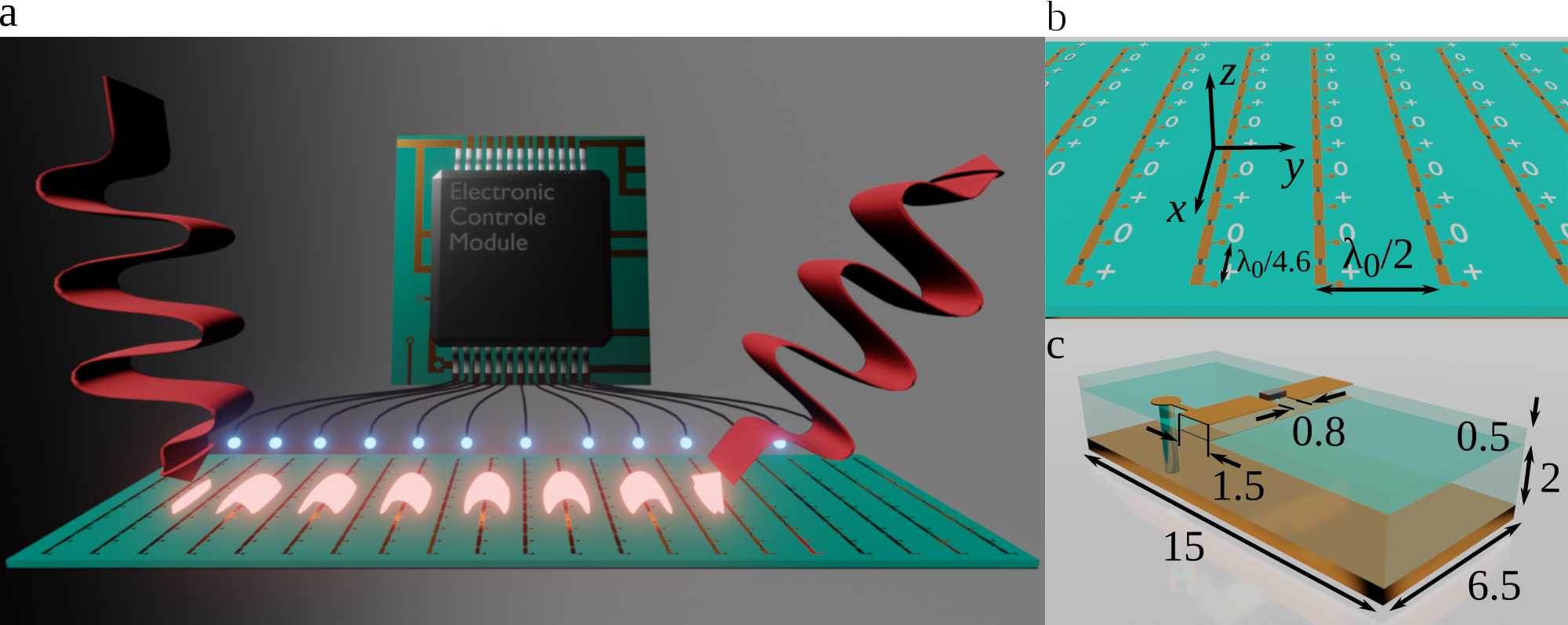}
	\caption{\label{fig:1} (a) An illustration of a  strongly non-local reconfigurable  sparse metasurface converting a normally incident wave into an anomalously reflected wave via an excited surface wave. 
	The surface wave propagates long the metasurface and manifests its strongly non-local response.  
	The electronic control module represents a multichannel DC voltage source run by a  Raspbery Pi nano-computer.
	(b) A close-up of a region of the sparse metasurface. (c) Schematics of an elementary varactor-loaded meta-atom composing the reconfigurable sparse metasurface, all the dimensions are given in millimeters. The bottom substrate (behind the ground plane) reserved for the bias network is not shown, see Supplementary Note 11.}
\end{figure*}
%

\section{Results}
\noindent\textbf{Theory}. 
Recently, it has been adopted that functionalities such as excitation of multiple beams do not require finely discretized metasurfaces and can be achieved with so-called metagratings~\cite{Alu2017_metagr,Epstein2017_mtg,Eleftheriades2018,Popov2018}. 
In contrast to metasurfaces, metagratings handle less elements and are not bounded by the local normal power flow conservation condition.
However, being periodic structures illuminated by a plane wave,  the functionality of metagratings is limited to the control of a finite number of Floquet modes.
Furthermore, a plane-wave illumination as in the case of dense metasurfaces significantly reduces the efficiency of the radiating system (a metagrating plus an external source) because of the spillover effect.
These two factors render the theoretical model of metagratings based on the Floquet periodicity condition impractical in real-life environment.
By dropping this constraint and considering finite-size structures straightaway,  we make a conceptual transition from meta\textit{gratings} to sparse metasurfaces.
In comparison to the former, sparse metasurfaces are able to arbitrarily control a \textit{continuous} spectrum of scattered waves, while respecting features of practical excitation sources, which never radiate a simple plane wave.
In what follows, we describe a theoretical approach of finite-size sparse metasurfaces.

A sparse metasurface cannot be described in terms of local reflection and/or transmission coefficients and continuous surface impedances, similarly to metagratings~\cite{Popov2018,Popov2019}.
Specifically, here we deal with a reflective sparse metasurface represented by a \textit{finite} array of $N$ loaded wires placed on top of a metal-backed dielectric substrate, as presented in Figure~\ref{fig:1}a.
From the microscopic perspective, a loaded wire is built up of subwavelength meta-atoms arranged in a line, as schematically shown in Figure~\ref{fig:1}b. Such system can be studied theoretically in accordance with the theoretical approach presented in~\cite{Popov2020}.
Namely, structured wires can be accurately approximated and modeled as uniform and having a deeply subwavelength effective radius $r_{eff}$.
The incident wave, which can be arbitrary and not only a plane wave, has its electric field polarized along the $x$-direction exciting polarization currents in the loaded wires.
Each current $I_q$ radiates the electric field $G_q(y,z)I_q$ with $G_q(y,z)$ being the Green's function corresponding to the $q$th wire at $y_q$ ($z_q=0$).
Then, the total scattered field $E_x(y,z)$ is a superposition of waves scattered by the wires and the incident wave reflected in the specular direction from the substrate $E_x^{(r)}(y,z)$ 
\begin{equation}\label{eq:field}
	E_x(y,z)=E_x^{(r)}(y,z)+\sum_{q=1}^N G_q(y,z)I_q.
\end{equation}
The scattering pattern is determined by values of the currents.
Meanwhile, each current $I_q$ is related via Ohm's law to a load-impedance density  $Z_q$ of the corresponding wire and the total electric field induced at its position 
\begin{equation}\label{eq:ohmslaw}
	Z_q I_q=E_x^{(exc)}(y_q,0)-Z^{(in)}_qI_q-\sum_{\substack{p=1\\ p\neq q}}^N Z^{(m)}_{qp}I_p,
\end{equation}
with $Z^{(m)}_{qp}=-G_p(y_q,0)$  and $Z^{(in)}_q=-G_q(y_q+r_{eff},0)$ being the mutual- and input-impedance densities.
A load-impedance density can be controlled by engineering the meta-atoms composing a wire.
Desirable scattering patterns can then be tailored by judiciously choosing $Z_q$.
For instance, the inverse scattering problem can be solved by maximizing power scattered by the metasurface in desirable directions in far-field or spots in near-field focusing with respect to load-impedance densities. The maximization procedure is implemented via particle swarm optimization~\cite{Kennedy_PSO}.
    
Since this study deals with a complex \textit{finite-size} system, there is no applicable analytical formula for a Green's function. 
Following Ref.~\cite{Popov2020}, we built a 2D simulation model using commercially available COMSOL Multiphysics FEM-based software to calculate Green's functions.
The model is represented by a finite-size grounded dielectric substrate placed in free-space.
To that end, an elementary source is placed at the positions of loaded wires, i.e. $y_q=q d$ ($q=1,2,...,N$, $z_q=0$). 
The schematics of the model are shown in Figure~S1 in the Supplementary Information.
Since there is no need to know Green's functions over the whole 2D plane, we extract the electric field created  by the source only at certain locations: (i) at the top of the dielectric substrate to construct the mutual-impedance matrix $Z_{qp}^{(m)}$ and find input-impedance densities $Z_{q}^{(in)}$, (ii) in the far field to show beam-forming performances, and (iii) in the desired focal plane to perform near-field focusing. 
Typical Green's functions are shown in Figure~S2 and S3 in the Supplementary Information.

\begin{figure*}[tb]
	\includegraphics[width=0.99\linewidth]{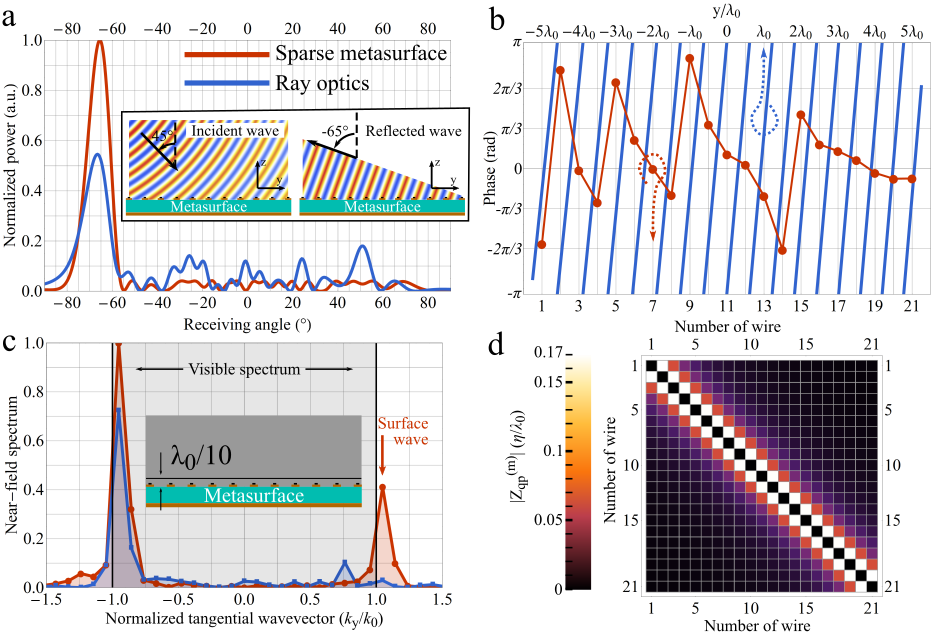}
	\caption{\label{fig:2} 
	(a) Scattering patterns from a flat sparse metasurface (red curve) and a phase gradient metasurface (blue curve) performing anomalous reflection at $-65^\circ$. The inset figure illustrates the incident (left) and reflected (right) waves. 
	(b) Corresponding phase of the local reflection coefficient along the phase gradient metasurface (blue curve) and 
	\textit{fictitious} phase of the local reflection coefficients corresponding to the sparse metasurface (red circles). 
	(c) Spectrum of the electric field above (at the distance $\lambda_0/10$) the sparse (red points) and gradient (blue points) metasurfaces. 
	(d) Plot of the matrix of mutual-impedance densities of the sparse metasurface. Parameters of the sparse metasurface are: $\varepsilon_s=2.2$, thickness of the PEC-backed substrate is $\lambda_0/11.3$ (PEC stands for the perfect electric conductor), the separation between two wires is $\lambda_0/2$, where $\lambda_0$ is the operating vacuum wavelength.}
\end{figure*}
%


\noindent\textbf{Numerical example}. 
Since the majority of reconfigurable metasurfaces operates under the ray optics approach, in what follows we compare the efficiency of a sparse metasurface to that of a phase-gradient metasurface by considering a simple illustrative example.
The sparse metasurface is represented by $21$ loaded wires equidistantly distributed along the top surface of a metal-backed dielectric substrate and the distance between two neighboring wires is $\lambda_0/2$ ($\lambda_0$ being the free-space central operating wavelength), as illustrated in Figure~\ref{fig:1}a,b.
In this example demonstrated by means of 2D full-wave simulations, the wires are modeled as hollow cylinders of radius $r_{eff}$ with an electric surface current density boundary condition of $E_x(y,z)/(2\pi r_{eff})/Z_q$.
To compare wavefront transformation performances between the sparse metasurface and a phase-gradient metasurface, we implement an extreme example of anomalous reflection of $-65^\circ$.
It is important to note that in conceptual studies, an incident wave is typically represented by a plane wave~\cite{Asadchy2016,Alu2016}. In this work, we deal with a cylindrical wave illumination.
Namely, the excitation is set as a TE-polarized cylindrical wave incident at $45^\circ$ and having the phase center $10\lambda_0$ away from the center of the metasurfaces, as shown in the inset of Figure~\ref{fig:2}a.
Such configuration makes the separation between the specular reflection and the anomalously reflected wave equal to $110^\circ$.
Both classical phase-gradient and sparse metasurfaces are considered to be purely reactive.
Ray optics approach suggests that the phase profile shown as the blue curve in Figure~\ref{fig:2}b should be established by the phase-gradient metasurface.
The red dot markers in Figure~\ref{fig:2}b correspond to the fictitious local reflection coefficient calculated for the sparse metasurface according to the retrieval procedure described in the Supplementary Note 3.
Since the red dots do not overlap with the blue curve, one can conclude that the approach of ray optics cannot be applied to design sparse metasurfaces.
Figure~\ref{fig:2}a  compares the scattering patterns created by  the sparse metasurface (red curve) and the phase-gradient metasurface (blue curve).
The classical metasurface exhibits strong spurious scattering in the far-field which makes the power scattered in the desired direction twice less than in case of the sparse metasurface. 
A key feature of sparse metasurfaces, allowing one to outperform gradient metasurfaces, is their strong non-locality. 
By extracting the scattered near-field at a distance $\lambda_0/10$ above the sparse metasurface and performing discrete Fourier transform~\cite{Oppenheim1989_DFT}, a strong peak manifesting a surface wave is observed outside the visible part of the near-field spectrum ($|k_y/k_0|>1$), as highlighted by the red trace in Figure~\ref{fig:2}c.
Although the interaction between elements of the sparse metasurface decreases with distance and well localized, as shown in Figure~\ref{fig:2}d, the surface wave propagates along the whole metasurface, representing a collective effect.
Conversely, the  spectrum of the scattered near-field above the gradient metasurface (blue line in Figure~\ref{fig:2}c) does not exhibit any signature of surface waves.


\noindent\textbf{Design}. By gaining control over load-impedance densities $Z_q$ of the wires, we are able to adjust the currents $I_q$ and manipulate the scattered field dynamically.
In order to realize real-time control over load-impedance densities at microwave frequencies, we use a radio-frequency varactor diode where the capacitance can be tuned by an applied bias voltage.
The schematic layout of the developed reconfigurable meta-atom is shown in Figure~\ref{fig:1}c. The design procedure was performed with the help of local periodic approximation recently established for periodic metagratings~\cite{Popov2019_LPA} since the conventional design methods of dense metasurfaces~\cite{Epstein:16} are not applicable for sparse metasurfaces.
Three principal parts of the elementary cell can be distinguished: a varactor loaded microstrip line (top), a microstrip line (middle), and a bias line (alongside). 
The microstrip lines together with the varactor diode form an equivalent parallel $RLC$ circuit. The middle microstrip line plays the role of an inductance that allows one to effectively decrease the minimal capacitance of the varactor diode and expand the dynamic range of accessible load-impedance densities $Z_q$  without increasing the density of meta-atoms. The latter is extremely important when one tends to decrease the Ohmic losses of varactor diodes. The bias voltage is applied to the varactor diode through the metallized via connecting the top and bottom layers.

\begin{figure*}[tb]
	\includegraphics[width=0.99\linewidth]{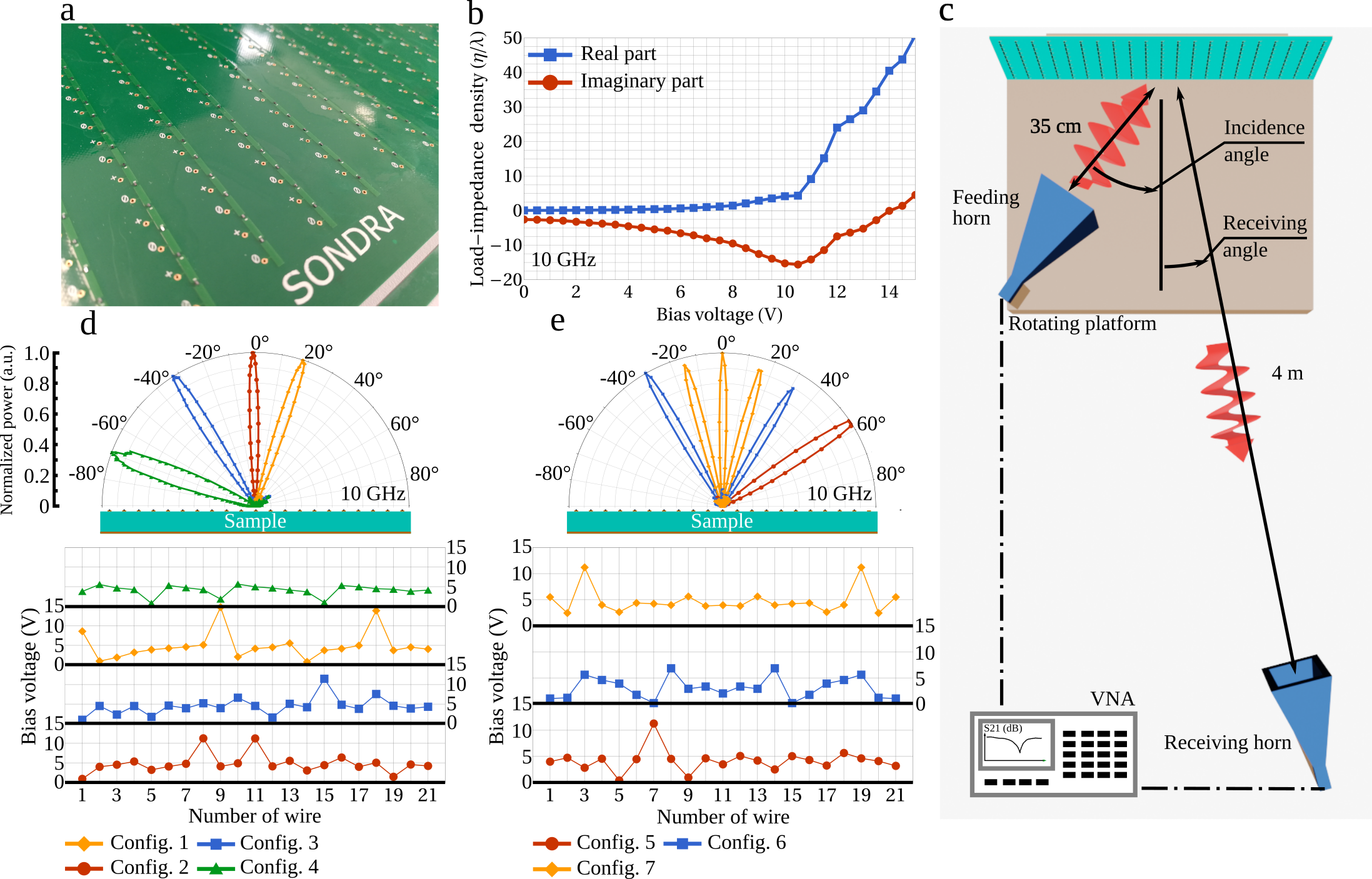}
	\caption{\label{fig:3} (a) A close-up photography of the fabricated sample. (b) Established dependence of the load-impedance density on applied bias voltage. (c) Schematic illustration of the experimental setup to measure the far-field patterns. (d),(e) Experimental examples of (d) beam-steering and (e) dynamic control of the number of beams, together with the bias voltages applied to each wire in the two different configurations.
	}
\end{figure*}

Similar to the theoretical example of anomalous reflection above, the experimental sample is composed of $21$ loaded wires, each represented by a chain of the designed meta-atoms, distributed along the surface of a metal-backed dielectric substrate, as shown in Figure~\ref{fig:1}a,b.
Individual control of each wire in the metasurface is implemented by a bias network appearing behind the metal-backed substrate and powered by a multichannel DC voltage source. 
Once the metasurface is fabricated using printed circuit board technique and surface-mount component soldering (Figure~\ref{fig:3}a), it is characterized to establish the dependence of the load-impedance density $Z_q$ upon the biasing voltage $V$. The characterization procedure is described in the Methods and Supplementary Information.
The measured load impedance density is plotted at $10$ GHz for different bias voltages in Figure~\ref{fig:3}b.

%

\noindent\textbf{Dynamic far-field beam-forming}.
The experimental setup used to demonstrate the dynamic far-field manipulation capabilities of the proposed metasurface is schematically illustrated in Figure~\ref{fig:3}c.
The sample and a feeding horn antenna are both placed on a rotating platform to be used as a transmitter and the radiation pattern is measured by a receiving horn antenna.
The feeding horn placed at $35$ cm ($\approx11.5\lambda_0$) away from the sample constitutes a cylindrical wave source, which makes it a very compact radiation system compared to generally exploited plane wave source. Other sources such as planar microstrip patch antenna can also be used in order to illuminate the sample and further reduce the profile of the system.
In the first set of measurements, the incidence angle is set to $45^\circ$ in concordance with the numerical example above, and the operating frequency is fixed to $10$ GHz.
Bias voltages for desired steering angles are found by means of the theoretical procedure described above and the experimental characterization of the sample detailed in Methods and Supplementary Note 4, and no further fine-tuning of voltages is required.
Figure~\ref{fig:3}d demonstrates dynamic beam steering from $20^\circ$ up to $-70^\circ$, 
the corresponding bias voltages are also shown.
The spurious scattering in the far-field region remains below 0.1, being well beyond limits of phase-gradient metasurfaces for large steering angles~\cite{Asadchy2016,Alu2016}.
As follows from the conclusions derived when analyzing Figure~\ref{fig:2}a,c, the efficient large-angle anomalous reflection serves as an indirect manifestation of a strong non-locality by the fabricated sparse reconfigurable metasurface.
Even though we deal with a \textit{sparse} metasurface, the employed design approach does not impose any restriction on the achievable beam-steering angle, which is in strong contrast with traditional periodic metasurface.

\begin{table*}[tb]
\caption{\label{tab:ff}
Summary of the performances of previously reported reconfigurable metasurfaces and our proposed sparse metasurface for application to beam-forming at microwave frequencies. The number of elements per $\lambda^2$ area is given for the corresponding operating frequency range.
$^*$ See Supplementary Information.
}
\resizebox{0.99\textwidth}{!}
{%
\begin{tabular}{|c|c|c|c|c|c|c|c|c|} 
\hline
\multirow{2}{*}{Reference} &  \textnumero$\,$ of elements& Frequency  & Instantaneous  & Incidence  & Max. reflection &  Level of & Total & Max. \textnumero \\ 
& per $\lambda^2$ area & range   & bandwidth & angle& angle  & side-lobes & absorption  & of  beams \\
\hline
D. Sievenpiper \textit{et al.}~\cite{Sievenpiper2003} & 74--44 & 3.5-4.5 GHz  & 8\% & $0^\circ$ & $40^\circ$ &  $\approx -10$ dB & $35\%$ for $0^\circ$ refl. angle &  1 \\  
\hline
H. Yang \textit{et al.}~\cite{yang2016programmable} & 5 & 11.1 GHz  & N/A & $0^\circ$ & $40^\circ$ &  $\approx -20$ dB & N/A &  3 \\  
\hline
B. Ratni \textit{et al.}~\cite{Ratni18} & 31--21 & 9-11 GHz & N/A & $0^\circ$ & $45^\circ$  & $\approx -10$ dB& $\approx 70\%$ & 1\\ 
\hline
\multirow{2}{*}{This work} & \multirow{2}{*}{13--8} & \multirow{2}{*}{8.5-10 GHz} & \multirow{2}{*}{5\%} & $45^\circ$  & $-70^\circ$  & -10 dB at 10 GHz & $^*51\%$ at 10 GHz & \multirow{2}{*}{$^* 6$}\\
\cline{5-8}
 & & & & $0^\circ$ & $^* 65^\circ$  & -12.5 dB at 9.5 GHz& $^*64\%$ at 9.5 GHz & \\
\hline
\end{tabular}
}
\end{table*}

The efficiency of beam-forming can be estimated more rigorously by calculating the scattering efficiency $\mathcal{E}_{sct}$ and the absorption $\mathcal A$ of the sample. The former is found as the power in the desired beam divided by the total reflected power
\begin{equation}
    \label{eq:sc_eff}
    \mathcal{E}_{sct}=\int_{\varphi_1}^{\varphi_2}P_s(\varphi)\textup{d}\varphi \bigg/
    \int_{-90^\circ}^{90^\circ}P_s(\varphi)\textup{d}\varphi,
\end{equation}
where $P_{s}(\varphi)$ is the scattered power measured in the angle $\varphi$, angles $\varphi_1$ and $\varphi_2$ confine the main lobe.
In Supplementary Note 6, we also compare the sample's directivity to  that of a uniform aperture of the same size, which serves as a standard reference. 
The scattering efficiency of the beam-steering functionality, displayed in Fig.~\ref{fig:3}d,  lies in the range of 72\% to 83\%.
The minimum efficiency corresponds to $-35^\circ$ steering and the maximum one to $-70^\circ$ steering.
This little bit counter-intuitive result can be explained on the one hand by the relatively large beamwidth ($\varphi_2-\varphi_1$) in the case of $-70^\circ$.
On the other hand, the level of side-lobes is the same for the beam-steering at $-35^\circ$ and $-70^\circ$.
It should also be noted that the scattering efficiency~\eqref{eq:sc_eff} of the non-periodic sparse metasurface cannot be directly compared to scattering efficiencies of metagratings reported in literature.
Indeed, when estimating experimental scattering efficiency of a metagrating, one generally normalizes the power scattered in a desired diffraction order to the total power scattered in all diffraction orders~\cite{Epstein2019_mtg_exp}.
While a fabricated metagrating always has a finite size and therefore creates a continuous angular spectrum of the scattered field,
such a procedure does not account for the power scattered in other directions apart from the diffraction orders represented by a finite set of Floquet modes. 
The absorbed energy by the sample can be estimated by normalizing the total power scattered from the sample to the one reflected from a metallic plate $P_{MP}(\varphi)$ of the same dimensions (see Supplementary Note 6 for details)
\begin{equation}
    \label{eq:absorption}
    \mathcal A=1-\int_{-90^\circ}^{90^\circ}P_s(\varphi)\textup{d}\varphi \bigg/
    \int_{-90^\circ}^{90^\circ}P_{MP}(\varphi)\textup{d}\varphi.
\end{equation}
In the examples demonstrated in Figure~\ref{fig:3}d, the absorption lies in the range from 40\% ($20^\circ$) to 51\% ($-70^\circ$).
The main source of the absorption are the varactor diodes which are relatively lossy at $10$ GHz.

More complex examples of beam-forming are illustrated in Figure~\ref{fig:3}e under $0^\circ$ incidence, which demonstrates the ability to dynamically control the number of radiated beams and also highlights the metasurface's agility with respect to the excitation. 
When applying successive sequence of pre-registered bias voltage profiles, one beam is firstly radiated at $60^\circ$, then two symmetrical beams at $-30^\circ$ and $30^\circ$, and finally three beams centered around $0^\circ$. 
Additional examples of beam-forming are demonstrated in Figure~S7-S9 in the Supplementary Information.
In all examples, the spurious scattering  does not exceed the $-10$ dB level. 
The high scattering efficiency of the sample is due to the rigorous theoretical model behind: we account for the mutual interactions between the elements, consider real-type excitation (non-planar incident wavefront in our specific case) and precisely solve the inverse scattering problem.
The instantaneous operating bandwidth of the sample also depends on the configuration and reaches approximately  $5$\%.
On the other hand, the reconfigurability mechanism allows the sample to demonstrate  frequency agility over a broad frequency range around centre frequency of $9.25$ GHz ($16$\%). 
The beamwidths are calculated assuming a level of side-lobes remaining below $-10$ dB.
The Supplementary Information provides additional details on the performances of the sample for beam-steering applications and on a much compact antenna system using a point source in close proximity of the sparse metasurface.

\begin{figure*}[tb]
	\includegraphics[width=0.99\linewidth]{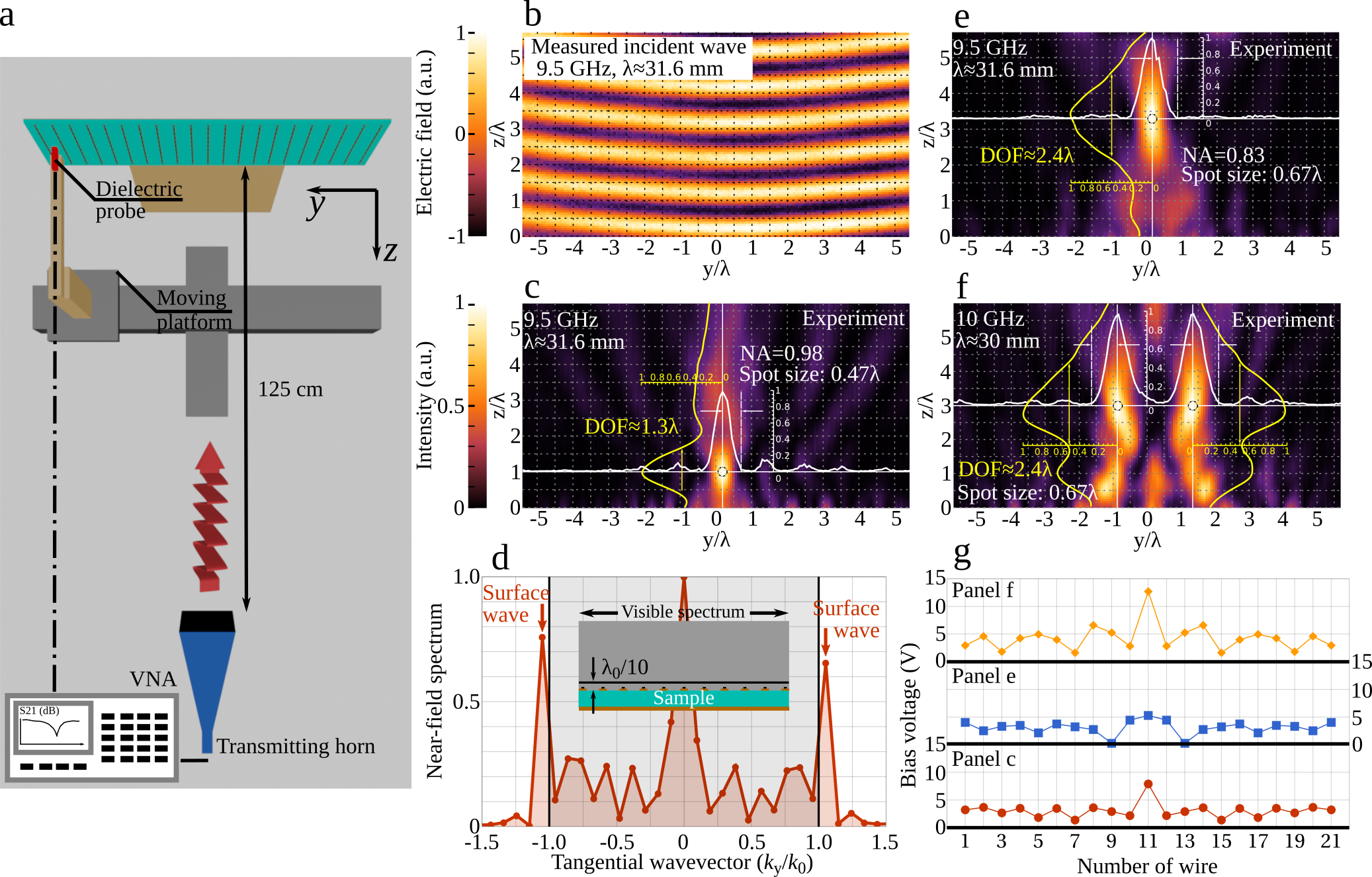}
	\caption{\label{fig:4} (a) Schematic illustration of the experimental setup to measure distribution of the electric field. (b) The measured electric field of the incident wave radiated by the transmitting horn antenna. (c), (e), (f) The measured normalized intensity of the scattered wave $|E_x(y,z)-E_x^{(inc)}(y,z)|^2$ for three different configurations of the reconfigurable sparse metasurface:  single focal spot at the distance $\lambda$ (c)  and $3\lambda$ (e), when frequency is $9.5$ GHz, (f) two focal spots at the distance $3\lambda$, when frequency is $10$ GHz. (d) Spectrum of the electric field along the sample measured at the distance $\lambda/10$ from the sample in the configuration presented in (c), showing surface waves propagating in opposite directions along the metasurface. 
	The panels (c), (e) and (f) represent processed experimental data, the processing technique is described in the Supplementary Information.
	(g) Bias voltages applied to each wire in the different configurations shown in panel (c), (e) and (f).}
\end{figure*}

Table~\ref{tab:ff} allows one to compare the efficiency and functionalities of the  reconfigurable sparse metasurface to traditional reconfigurable metasurfaces.
From this comparison, it is clearly observed that the implemented mechanism of strong electromagnetic non-locality allows the reconfigurable sparse metasurface to outperform traditional designs, reach larger steering angles and effectively excite multiple beams.
Furthermore, the levels of side-lobes and absorption are lower or similar to those of traditional metasurfaces demonstrating more modest functionality.

\noindent\textbf{Adaptive near-field focusing}. 
To further demonstrate the vast capabilities of the approach in manipulating fields, we study the flexibility and efficiency of the proposed reconfigurable metasurface for dynamic shaping of the near-field.
Another experimental setup is mounted to measure the  distribution of the electric field in front of the sample with a moving dielectric probe as sketched in Figure~\ref{fig:4}a. The EFS-105-12 fiber optic active antenna is used as field-sensing probe to measure both the amplitude and the phase of the electric field. It is composed of a purely dielectric head of 6.6 x 6.6 mm$^2$ cross-section with negligible field perturbation~\cite{probe}.
The sample is illuminated at normal incidence by a cylindrical wave radiated by a horn antenna, as illustrated by the corresponding measured electric field profile in Figure~\ref{fig:4}b. The processed experimental data are presented in Figures~\ref{fig:4}c, e and f. The processing procedure is described and compared to raw data in the Supplementary Note 7.
In the first sequence of voltage settings, the sample acts as a high numerical aperture (NA) diffractive lens and focuses the incident wave at the distance $\lambda$ as shown in Figure~\ref{fig:4}c.
With the physical aperture of the sample being $21\lambda_0/2=315$ mm and the focal length of $\lambda\approx31.6$ mm, the NA is approximately $0.98$.
The experimentally measured spot size being $0.47\lambda$ (corresponding FWHM is $0.43\lambda$), which is smaller (for the given NA) than the size of Airy spot $0.61\lambda/\textup{NA}$ establishing the diffraction limit~\cite{abbe1873beitrage,novotny2012principles}, demonstrates subdiffraction focusing.
Meanwhile, the minimum spot size  of a subdiffraction focal spot is found to be $0.38\lambda/\textup{NA}$~\cite{Qiu2018_diff_lens}.
The  depth of focus (DOF), characterized as full width of the focal spot in the longitudinal direction at half maximum, is experimentally measured as $1.3\lambda$, which agrees very well with the theoretical estimation $\textup{DOF}=\lambda/(1-\sqrt{1-\textup{NA}^2})\approx 1.25\lambda$~\cite{Qiu2018_diff_lens}.
The imaging and total efficiencies in this configuration are $68\%$ and $21\%$, respectively.
The level of secondary lobes is -8.5 dB.
The imaging (total) efficiency is calculated as the reflected power in the focal spot divided over the total reflected (incident) power in the focal plane
\begin{equation}\label{eq:eff_img}
\mathcal E_{img}=\int_{y_1}^{y_2}|E_x(y,z_f)-E^{(inc)}_x(y,z_f)|^2\textup{d}y\bigg/\int_{y_{min}}^{y_{max}}|E_x(y,z_f)-E^{(inc)}_x(y,z_f)|^2\textup{d}y,    
\end{equation}
\begin{equation}\label{eq:eff_tot}
\mathcal E_{tot}=\int_{y_1}^{y_2}|E_x(y,z_f)-E^{(inc)}_x(y,z_f)|^2\textup{d}y\bigg/\int_{y_{min}}^{y_{max}}|E^{(inc)}_x(y,z_f)|^2\textup{d}y,    
\end{equation}
where the coordinates $y_1$ and $y_2$ refer to the lateral limits of the focal spot and the coordinates $y_{min}$ and $y_{max}$ correspond to the lateral boundaries of the scanned region.

\begin{table*}[tb]
\caption{\label{tab:nf}
Summary of the performances of previously reported reconfigurable metasurfaces in application to focusing at microwave frequencies. $^*$See Supplementary Information.
}
\resizebox{0.99\textwidth}{!}
{%
\begin{tabular}{|c|c|c|c|c|c|c|c|} 
\hline
\multirow{2}{*}{Reference} & \textnumero$\,$ of elements & \multirow{2}{*}{Frequency} & Focal  & \multirow{2}{*}{FWHM ($\lambda$)} &  Imaging & Total & Max. \textnumero \\ 
& per $\lambda^2$ area & & length ($\lambda$) & & efficiency (\%) &  efficiency (\%) & of spots \\
\hline
W. Zhu \textit{et al.}~\cite{Zheludev2015} & 14 & 16 GHz & 5.1, 10, 15.2 & 2.1, 2.2, 2.5 & N/A & 10 & 1\\
\hline
H.-X. Xu \textit{et al.}~\cite{zhou2016focusing} & 21 & 5.5 GHz & 0.83,1.38,1.65,2.20 & --,$0.61$,$0.63$,-- & N/A & N/A & 1 \\  
\hline
K. Chen \textit{et al.}~\cite{chen2017reconfigurable} & 27 & 6.9 GHz & 1, 1.5, 2, 4 & N/A & N/A & 36 & 2 \\  
\hline
B. Ratni \textit{et al.}~\cite{Ratni2020} & 31 & 9 GHz & 1.5, 2.25, 3 & 0.53, 0.6, 0.69  & 52, 56, 60 & 33, 40, 42 & 2\\ 
\hline
\multirow{2}{*}{This work} & \multirow{2}{*}{8 at 10 GHz} & 9.5 GHz& 1, 3 (Figs.~\ref{fig:4}c and e)   & 0.43, 0.6 & 68, 78 & 21, 30& \multirow{2}{*}{$^*5$}\\
\cline{3-7}
 & &  10 GHz&  3 (Fig.~\ref{fig:4}f)  &  0.6 & 88 &  57 & \\
\hline
\end{tabular}
}
\end{table*}

Extracting the amplitude and phase of the electric field in the proximity of the sample allows us to \textit{directly} observe surface waves manifesting strong non-locality.
Performing  discrete Fourier transform~\cite{Oppenheim1989_DFT} on the measured electric field at the distance $\lambda/10$ from the metasurface reveals two strong peaks outside the visible range ($k_y/k_0>1$), as shown in Figure~\ref{fig:4}d. 
The peaks correspond to two surface waves  propagating along the sample in opposite directions.
The surface waves contribute to the aperture field and through the aperture field they are responsible for breaking the diffraction limit. 
However, being evanescent, the surface waves do not constitute the focal spot at $\lambda$ distance from the sample.
Indeed, the spectrum of the field in the focal plane consists in only propagating waves as shown in  Figure~S19 in the Supplementary Information.
On the other hand, if one moves the focal spot closer to the metasurface,  surface waves can be used to perform super-focusing and achieve the spot size much less than that of the subdiffraction limit  ($0.38\lambda$, $\textup{NA}\approx 1$). 
As demonstrated in the Supplementary Information,  to that end a denser, but still sparse, metasurface is required~\cite{Popov2019}.
In Figure~S21-S24 an example of a sparse metasurface with the inter-wire distance of $\lambda_0/4$ demonstrates focusing of the incident wave at $\lambda/10$ focal length to the spot size of $0.15\lambda$.
In strong contrast to the phenomenon of super-oscillations~\cite{Qiu2018_diff_lens}, which also allows to overcome the subdiffraction limit, the level of side-lobes remains very low due to the surface waves contributing significantly to the formation of the focal spot as revealed in the Supplementary Information.

By changing the applied voltage sequence and thus the load-impedance densities, one can move the focal spot further away from the sample and change the NA of the lens (from $0.98$ in Figure~\ref{fig:4}c to 
$0.83$
in Figure~\ref{fig:4}e).
The spot size of 
$0.67\lambda$ 
(FWHM is $0.6\lambda$) is still smaller than the size  $0.74\lambda$ of the corresponding Airy spot.
Multiple subdiffraction focal spots can also be created and independently controlled (Figure~\ref{fig:4}f and  Figure~S16-S18 in the Supplementary Information), demonstrating the high efficiency and manipulation flexibility of the real-time reconfigurable metasurface.
In Figure~\ref{fig:4}e,f, the measured DOF and its corresponding theoretical estimation is, respectively, $2.4\lambda$ and  $2.26\lambda$.
At 9.5 GHz, for a single focal spot at $3\lambda$ distance (Figure~\ref{fig:4}e), the imaging and total efficiencies constitute, respectively, $78\%$ and $30\%$.
The increase of the total efficiency to $30\%$ (in comparison to $21\%$ for Figure~\ref{fig:4}c) is due to the lower level of side-lobes (-12 dB) and the higher imaging efficiency ($78\%$ vs. $68\%$).
The total (imaging) efficiency increases to $57\%$ ($88\%$) for two focal spots at 10 GHz displayed in Figure~\ref{fig:4}f.  
The difference between the total efficiencies at 9.5 and 10 GHz can be explained by different resonant behavior of the load-impedance density at these two frequencies, as it can be compared from Figures~S6a and b.
Indeed, when creating a single focal spot at 10 GHz (shown in Figure~S18c in the Supplementary Information), the total efficiency also increases to $45\%$ and the imaging efficiency reaches $86\%$.

Table~\ref{tab:nf} compares the focusing performances of the tested sample with those of conventional reconfigurable metasurfaces reported previously in literature.
The comparison shows that the strong non-locality implemented in our reconfigurable metasurface allows achieving superior performances in terms of spot size and imaging efficiency for multiple functionalities.
Furthermore, the total efficiency of the fabricated sparse metasurface can exceed those of traditional designs due to lower density of elements and operation outside resonance, where mainly capacitive response is required.

\section{Discussion and conclusion}
In this work we have demonstrated implementation of the mechanism of strong electromagnetic non-locality in a reconfigurable sparse metasurface.
The design procedure of such a metasurface has been described in details: from the theoretical model based on a numerical calculation of a Green's function to the microscopic design of a unit cell incorporating a tunable element.
It can be especially useful for implementation of reconfigurable metasurfaces at optical and visible frequencies where they suffer a lot from efficiency problems.
Indeed, by reducing the density of meta-atoms and thus, complexity of the design and fabrication of a reconfigurable metasurface,
one  is actually able to outperform conventional metasurfaces, which often require more sophisticated design.
Furthermore, the generality of the theoretical approach allows one to design conformal reconfigurable metasurfaces as well as planar ones.

As a direct endorsement of the proposed approach, a reconfigurable sparse metasurface loaded with controllable varactor diodes has been proposed to efficiently and dynamically manipulate fields in both near-field and far-field regions with an arbitrary incident wave illumination.
Fundamentally, we have experimentally demonstrated direct and indirect evidences of surface waves propagating along the reconfigurable sparse metasurface manifesting its strongly non-local response.
By comparing experimental performances of the fabricated sample to others reported previously in literature, we have demonstrated that the dynamic strongly non-local response allows to outperform conventional designs of reconfigurable metasurfaces in a broad range of functionalities both in near-field and far-field regions.
As such, subdiffraction focusing, anomalous reflection at extreme angles, dynamic control of multiple beams and focal spots have been demonstrated.
The ability of the proposed sparse metasurface to dynamically manipulate near- and far-fields in real-time paves the way to applications in reconfigurable devices such as lenses, antennas and imaging systems with superior performances.


\section*{Methods}

\noindent\textbf{Sample fabrication}. The sample was fabricated by means of the conventional printed circuit board (PCB) technique and represents a six-layer sample with the bottom layer reserved for the bias network of the varactor diodes. 
Three woven-glass dielectric substrates F4BM220 (relative permittivity is $2.2$, loss tangent is $0.001$) were used: one of $2$ mm thickness and two of $0.5$ mm thickness. 
Thickness of the copper cladding is $35$ $\mu$m.
Selected varactor diode (MAVR-011020-1411) has capacitance varying in the range from $0.045$ pF to $0.25$ pF. 
Metallic vias of $0.25$ mm radius are used to apply bias voltage to varactors. 

\noindent\textbf{Electronic control module}.  
In order to control the fabricated sample, an electronic control module was developed upstream. 
It consists of a Raspberry Pi 3 B+, an ARM processor single board nano-computer, and a stack of two electronic cards integrating 16 independent DC voltage outputs each.
The nano-computer is used as an operating unit for the two electronic cards and allows one to control the DC voltage of each of the 32 outputs from 0 to 30V.
To facilitate the control, a human-machine interface in Python was developed.
On the electronic side, each  card is supplied with a fixed voltage of 30 V and operational amplifiers are used in a comparator assembly to control the voltage of each output.
Each output is therefore controlled via the  nano-computer using the SPI communication protocol.

\noindent\textbf{Characterization}. The aim of the experimental characterization of the sample is to establish the dependence between the applied bias voltage and the load-impedance density of the loaded wire comprising the sparse metasurface. 
A schematics and a photography of the experimental setup are shown respectively in Figure~S5a,b in the Supplementary Information.
We measure the frequency dependence of the complex amplitude of the specularly reflected wave from the sample as a function of the applied bias voltage.
The same bias voltage is applied to all varactor diodes.
The complex amplitude is normalized to the one measured from a metallic plate of the same size and thickness as the sample.
The absolute value of the reflection coefficient  is plotted in Figure~S5c in the Supplementary Information.
Then, we correlate the experimental frequency dependencies  with the ones  obtained via 3D full-wave numerical simulations  of a unit cell (Figure~\ref{fig:1}c) with applied periodic boundary conditions, see Figure~S5d in the Supplementary Information. 
In the simulations the varactor diode is modeled as a lumped RC-circuit with the impedance $Z_{v}=R_v-j/(\omega C_v)$, where $R_v$ is the series resistance and $C_v$ is the capacitance.
It is assumed that neither $R_v$ nor $C_v$ depend on the frequency.
By comparing the positions of the experimental and theoretical resonances (Figure~S5e in the Supplementary Information), we obtain the relation between the bias voltage and the varactor’s capacitance presented in Figure~S5f in the Supplementary Information. 
Eventually, with $C_v=C_v(V)$ known, we are able to find $Z_q=Z_q(V)$.


\section*{Data availability}
The data that support the findings of this study are available from the authors on reasonable request.

\section*{Author contributions}
V.P. and F.B. conceived the concept, V.P. developed the theory and the design methodology, performed numerical simulations, V.P., F.B. and S.N.B. carried out the design of the experimental samples, F.B. and S.N.B. oversaw the manufacture of the samples, B.R. participated in development of the multichannel voltage source used to power the reconfigurable sample, V.P. and B.R. performed experimental measurements.
All the authors contributed to the interpretation of results and participated in the preparation of the manuscript.

\section*{Additional information}
Supplementary Information accompanies this paper at [URL provided by the publisher].

\section*{Conflict of interests}
The authors declare no competing financial interests.

\bibliography{bib}

\providecommand{\noopsort}[1]{}\providecommand{\singleletter}[1]{#1}%
\begin{thebibliography}{66}%
\makeatletter
\providecommand \@ifxundefined [1]{%
 \@ifx{#1\undefined}
}%
\providecommand \@ifnum [1]{%
 \ifnum #1\expandafter \@firstoftwo
 \else \expandafter \@secondoftwo
 \fi
}%
\providecommand \@ifx [1]{%
 \ifx #1\expandafter \@firstoftwo
 \else \expandafter \@secondoftwo
 \fi
}%
\providecommand \natexlab [1]{#1}%
\providecommand \enquote  [1]{``#1''}%
\providecommand \bibnamefont  [1]{#1}%
\providecommand \bibfnamefont [1]{#1}%
\providecommand \citenamefont [1]{#1}%
\providecommand \href@noop [0]{\@secondoftwo}%
\providecommand \href [0]{\begingroup \@sanitize@url \@href}%
\providecommand \@href[1]{\@@startlink{#1}\@@href}%
\providecommand \@@href[1]{\endgroup#1\@@endlink}%
\providecommand \@sanitize@url [0]{\catcode `\\12\catcode `\$12\catcode
  `\&12\catcode `\#12\catcode `\^12\catcode `\_12\catcode `\%12\relax}%
\providecommand \@@startlink[1]{}%
\providecommand \@@endlink[0]{}%
\providecommand \url  [0]{\begingroup\@sanitize@url \@url }%
\providecommand \@url [1]{\endgroup\@href {#1}{\urlprefix }}%
\providecommand \urlprefix  [0]{URL }%
\providecommand \Eprint [0]{\href }%
\providecommand \doibase [0]{http://dx.doi.org/}%
\providecommand \selectlanguage [0]{\@gobble}%
\providecommand \bibinfo  [0]{\@secondoftwo}%
\providecommand \bibfield  [0]{\@secondoftwo}%
\providecommand \translation [1]{[#1]}%
\providecommand \BibitemOpen [0]{}%
\providecommand \bibitemStop [0]{}%
\providecommand \bibitemNoStop [0]{.\EOS\space}%
\providecommand \EOS [0]{\spacefactor3000\relax}%
\providecommand \BibitemShut  [1]{\csname bibitem#1\endcsname}%
\let\auto@bib@innerbib\@empty
\bibitem [{\citenamefont {{Sievenpiper}}\ \emph {et~al.}(2003)\citenamefont
  {{Sievenpiper}}, \citenamefont {{Schaffner}}, \citenamefont {{Song}},
  \citenamefont {{Loo}},\ and\ \citenamefont {{Tangonan}}}]{Sievenpiper2003}%
  \BibitemOpen
  \bibfield  {author} {\bibinfo {author} {\bibfnamefont {D.~F.}\ \bibnamefont
  {{Sievenpiper}}}, \bibinfo {author} {\bibfnamefont {J.~H.}\ \bibnamefont
  {{Schaffner}}}, \bibinfo {author} {\bibfnamefont {H.~J.}\ \bibnamefont
  {{Song}}}, \bibinfo {author} {\bibfnamefont {R.~Y.}\ \bibnamefont {{Loo}}}, \
  and\ \bibinfo {author} {\bibfnamefont {G.}~\bibnamefont {{Tangonan}}},\
  }\bibfield  {title} {\enquote {\bibinfo {title} {Two-dimensional beam
  steering using an electrically tunable impedance surface},}\ }\href@noop {}
  {\bibfield  {journal} {\bibinfo  {journal} {IEEE Transactions on Antennas and
  Propagation}\ }\textbf {\bibinfo {volume} {51}},\ \bibinfo {pages}
  {2713--2722} (\bibinfo {year} {2003})}\BibitemShut {NoStop}%
\bibitem [{\citenamefont {{Lau}}\ and\ \citenamefont
  {{Hum}}(2012)}]{Hum2012_transmitarray}%
  \BibitemOpen
  \bibfield  {author} {\bibinfo {author} {\bibfnamefont {J.~Y.}\ \bibnamefont
  {{Lau}}}\ and\ \bibinfo {author} {\bibfnamefont {S.~V.}\ \bibnamefont
  {{Hum}}},\ }\bibfield  {title} {\enquote {\bibinfo {title} {Reconfigurable
  transmitarray design approaches for beamforming applications},}\ }\href
  {\doibase 10.1109/TAP.2012.2213054} {\bibfield  {journal} {\bibinfo
  {journal} {IEEE Transactions on Antennas and Propagation}\ }\textbf {\bibinfo
  {volume} {60}},\ \bibinfo {pages} {5679--5689} (\bibinfo {year}
  {2012})}\BibitemShut {NoStop}%
\bibitem [{\citenamefont {{Clemente}}\ \emph {et~al.}(2013)\citenamefont
  {{Clemente}}, \citenamefont {{Dussopt}}, \citenamefont {{Sauleau}},
  \citenamefont {{Potier}},\ and\ \citenamefont
  {{Pouliguen}}}]{Clemente2013_coding_transmitarray}%
  \BibitemOpen
  \bibfield  {author} {\bibinfo {author} {\bibfnamefont {A.}~\bibnamefont
  {{Clemente}}}, \bibinfo {author} {\bibfnamefont {L.}~\bibnamefont
  {{Dussopt}}}, \bibinfo {author} {\bibfnamefont {R.}~\bibnamefont
  {{Sauleau}}}, \bibinfo {author} {\bibfnamefont {P.}~\bibnamefont {{Potier}}},
  \ and\ \bibinfo {author} {\bibfnamefont {P.}~\bibnamefont {{Pouliguen}}},\
  }\bibfield  {title} {\enquote {\bibinfo {title} {Wideband 400-element
  electronically reconfigurable transmitarray in x band},}\ }\href {\doibase
  10.1109/TAP.2013.2271493} {\bibfield  {journal} {\bibinfo  {journal} {IEEE
  Transactions on Antennas and Propagation}\ }\textbf {\bibinfo {volume}
  {61}},\ \bibinfo {pages} {5017--5027} (\bibinfo {year} {2013})}\BibitemShut
  {NoStop}%
\bibitem [{\citenamefont {Zhu}\ \emph {et~al.}(2014)\citenamefont {Zhu},
  \citenamefont {Chen}, \citenamefont {Jia}, \citenamefont {Sun}, \citenamefont
  {Zhao}, \citenamefont {Jiang},\ and\ \citenamefont {Feng}}]{zhu2014dynamic}%
  \BibitemOpen
  \bibfield  {author} {\bibinfo {author} {\bibfnamefont {B.~O.}\ \bibnamefont
  {Zhu}}, \bibinfo {author} {\bibfnamefont {K.}~\bibnamefont {Chen}}, \bibinfo
  {author} {\bibfnamefont {N.}~\bibnamefont {Jia}}, \bibinfo {author}
  {\bibfnamefont {L.}~\bibnamefont {Sun}}, \bibinfo {author} {\bibfnamefont
  {J.}~\bibnamefont {Zhao}}, \bibinfo {author} {\bibfnamefont {T.}~\bibnamefont
  {Jiang}}, \ and\ \bibinfo {author} {\bibfnamefont {Y.}~\bibnamefont {Feng}},\
  }\bibfield  {title} {\enquote {\bibinfo {title} {Dynamic control of
  electromagnetic wave propagation with the equivalent principle inspired
  tunable metasurface},}\ }\href@noop {} {\bibfield  {journal} {\bibinfo
  {journal} {Scientific reports}\ }\textbf {\bibinfo {volume} {4}},\ \bibinfo
  {pages} {1--7} (\bibinfo {year} {2014})}\BibitemShut {NoStop}%
\bibitem [{\citenamefont {Cui}\ \emph {et~al.}(2014)\citenamefont {Cui},
  \citenamefont {Qi}, \citenamefont {Wan}, \citenamefont {Zhao},\ and\
  \citenamefont {Cheng}}]{cui2014coding}%
  \BibitemOpen
  \bibfield  {author} {\bibinfo {author} {\bibfnamefont {T.~J.}\ \bibnamefont
  {Cui}}, \bibinfo {author} {\bibfnamefont {M.~Q.}\ \bibnamefont {Qi}},
  \bibinfo {author} {\bibfnamefont {X.}~\bibnamefont {Wan}}, \bibinfo {author}
  {\bibfnamefont {J.}~\bibnamefont {Zhao}}, \ and\ \bibinfo {author}
  {\bibfnamefont {Q.}~\bibnamefont {Cheng}},\ }\bibfield  {title} {\enquote
  {\bibinfo {title} {Coding metamaterials, digital metamaterials and
  programmable metamaterials},}\ }\href@noop {} {\bibfield  {journal} {\bibinfo
   {journal} {Light: Science \& Applications}\ }\textbf {\bibinfo {volume}
  {3}},\ \bibinfo {pages} {e218} (\bibinfo {year} {2014})}\BibitemShut
  {NoStop}%
\bibitem [{\citenamefont {Della~Giovampaola}\ and\ \citenamefont
  {Engheta}(2014)}]{della2014digital}%
  \BibitemOpen
  \bibfield  {author} {\bibinfo {author} {\bibfnamefont {C.}~\bibnamefont
  {Della~Giovampaola}}\ and\ \bibinfo {author} {\bibfnamefont {N.}~\bibnamefont
  {Engheta}},\ }\bibfield  {title} {\enquote {\bibinfo {title} {Digital
  metamaterials},}\ }\href@noop {} {\bibfield  {journal} {\bibinfo  {journal}
  {Nature materials}\ }\textbf {\bibinfo {volume} {13}},\ \bibinfo {pages}
  {1115} (\bibinfo {year} {2014})}\BibitemShut {NoStop}%
\bibitem [{\citenamefont {Kaina}\ \emph {et~al.}(2014)\citenamefont {Kaina},
  \citenamefont {Dupr{\'e}}, \citenamefont {Lerosey},\ and\ \citenamefont
  {Fink}}]{kaina2014shaping}%
  \BibitemOpen
  \bibfield  {author} {\bibinfo {author} {\bibfnamefont {N.}~\bibnamefont
  {Kaina}}, \bibinfo {author} {\bibfnamefont {M.}~\bibnamefont {Dupr{\'e}}},
  \bibinfo {author} {\bibfnamefont {G.}~\bibnamefont {Lerosey}}, \ and\
  \bibinfo {author} {\bibfnamefont {M.}~\bibnamefont {Fink}},\ }\bibfield
  {title} {\enquote {\bibinfo {title} {Shaping complex microwave fields in
  reverberating media with binary tunable metasurfaces},}\ }\href@noop {}
  {\bibfield  {journal} {\bibinfo  {journal} {Scientific reports}\ }\textbf
  {\bibinfo {volume} {4}},\ \bibinfo {pages} {1--8} (\bibinfo {year}
  {2014})}\BibitemShut {NoStop}%
\bibitem [{\citenamefont {Zhu}\ \emph {et~al.}(2015)\citenamefont {Zhu},
  \citenamefont {Song}, \citenamefont {Yan}, \citenamefont {Zhang},
  \citenamefont {Wu}, \citenamefont {Chin}, \citenamefont {Cai}, \citenamefont
  {Tsai}, \citenamefont {Shen}, \citenamefont {Deng}, \citenamefont {Ting},
  \citenamefont {Gu}, \citenamefont {Lo}, \citenamefont {Kwong}, \citenamefont
  {Yang}, \citenamefont {Huang}, \citenamefont {Liu},\ and\ \citenamefont
  {Zheludev}}]{Zheludev2015}%
  \BibitemOpen
  \bibfield  {author} {\bibinfo {author} {\bibfnamefont {W.}~\bibnamefont
  {Zhu}}, \bibinfo {author} {\bibfnamefont {Q.}~\bibnamefont {Song}}, \bibinfo
  {author} {\bibfnamefont {L.}~\bibnamefont {Yan}}, \bibinfo {author}
  {\bibfnamefont {W.}~\bibnamefont {Zhang}}, \bibinfo {author} {\bibfnamefont
  {P-C.}\ \bibnamefont {Wu}}, \bibinfo {author} {\bibfnamefont {L.~K.}\
  \bibnamefont {Chin}}, \bibinfo {author} {\bibfnamefont {H.}~\bibnamefont
  {Cai}}, \bibinfo {author} {\bibfnamefont {D.~P.}\ \bibnamefont {Tsai}},
  \bibinfo {author} {\bibfnamefont {Z.~X.}\ \bibnamefont {Shen}}, \bibinfo
  {author} {\bibfnamefont {T.~W.}\ \bibnamefont {Deng}}, \bibinfo {author}
  {\bibfnamefont {S.~K.}\ \bibnamefont {Ting}}, \bibinfo {author}
  {\bibfnamefont {Y.}~\bibnamefont {Gu}}, \bibinfo {author} {\bibfnamefont
  {G.~Q.}\ \bibnamefont {Lo}}, \bibinfo {author} {\bibfnamefont {D.~L.}\
  \bibnamefont {Kwong}}, \bibinfo {author} {\bibfnamefont {Z.~C.}\ \bibnamefont
  {Yang}}, \bibinfo {author} {\bibfnamefont {R.}~\bibnamefont {Huang}},
  \bibinfo {author} {\bibfnamefont {A-Q.}\ \bibnamefont {Liu}}, \ and\ \bibinfo
  {author} {\bibfnamefont {N.}~\bibnamefont {Zheludev}},\ }\bibfield  {title}
  {\enquote {\bibinfo {title} {A flat lens with tunable phase gradient by using
  random access reconfigurable metamaterial},}\ }\href {\doibase
  10.1002/adma.201501943} {\bibfield  {journal} {\bibinfo  {journal} {Advanced
  Materials}\ }\textbf {\bibinfo {volume} {27}},\ \bibinfo {pages} {4739--4743}
  (\bibinfo {year} {2015})}\BibitemShut {NoStop}%
\bibitem [{\citenamefont {Dupr\'e}\ \emph {et~al.}(2015)\citenamefont
  {Dupr\'e}, \citenamefont {del Hougne}, \citenamefont {Fink}, \citenamefont
  {Lemoult},\ and\ \citenamefont {Lerosey}}]{Hougne2015}%
  \BibitemOpen
  \bibfield  {author} {\bibinfo {author} {\bibfnamefont {M.}~\bibnamefont
  {Dupr\'e}}, \bibinfo {author} {\bibfnamefont {P.}~\bibnamefont {del Hougne}},
  \bibinfo {author} {\bibfnamefont {M.}~\bibnamefont {Fink}}, \bibinfo {author}
  {\bibfnamefont {F.}~\bibnamefont {Lemoult}}, \ and\ \bibinfo {author}
  {\bibfnamefont {G.}~\bibnamefont {Lerosey}},\ }\bibfield  {title} {\enquote
  {\bibinfo {title} {Wave-field shaping in cavities: Waves trapped in a box
  with controllable boundaries},}\ }\href {\doibase
  10.1103/PhysRevLett.115.017701} {\bibfield  {journal} {\bibinfo  {journal}
  {Phys. Rev. Lett.}\ }\textbf {\bibinfo {volume} {115}},\ \bibinfo {pages}
  {017701} (\bibinfo {year} {2015})}\BibitemShut {NoStop}%
\bibitem [{\citenamefont {Xu}\ \emph {et~al.}(2016{\natexlab{a}})\citenamefont
  {Xu}, \citenamefont {Sun}, \citenamefont {Tang}, \citenamefont {Ma},
  \citenamefont {He}, \citenamefont {Wang}, \citenamefont {Cai}, \citenamefont
  {Li},\ and\ \citenamefont {Zhou}}]{xu2016dynamical}%
  \BibitemOpen
  \bibfield  {author} {\bibinfo {author} {\bibfnamefont {H.-X.}\ \bibnamefont
  {Xu}}, \bibinfo {author} {\bibfnamefont {S.}~\bibnamefont {Sun}}, \bibinfo
  {author} {\bibfnamefont {S.}~\bibnamefont {Tang}}, \bibinfo {author}
  {\bibfnamefont {S.}~\bibnamefont {Ma}}, \bibinfo {author} {\bibfnamefont
  {Q.}~\bibnamefont {He}}, \bibinfo {author} {\bibfnamefont {G.-M.}\
  \bibnamefont {Wang}}, \bibinfo {author} {\bibfnamefont {T.}~\bibnamefont
  {Cai}}, \bibinfo {author} {\bibfnamefont {H.-P.}\ \bibnamefont {Li}}, \ and\
  \bibinfo {author} {\bibfnamefont {L.}~\bibnamefont {Zhou}},\ }\bibfield
  {title} {\enquote {\bibinfo {title} {Dynamical control on helicity of
  electromagnetic waves by tunable metasurfaces},}\ }\href@noop {} {\bibfield
  {journal} {\bibinfo  {journal} {Scientific reports}\ }\textbf {\bibinfo
  {volume} {6}},\ \bibinfo {pages} {1--10} (\bibinfo {year}
  {2016}{\natexlab{a}})}\BibitemShut {NoStop}%
\bibitem [{\citenamefont {Yang}\ \emph {et~al.}(2016)\citenamefont {Yang},
  \citenamefont {Cao}, \citenamefont {Yang}, \citenamefont {Gao}, \citenamefont
  {Xu}, \citenamefont {Li}, \citenamefont {Chen}, \citenamefont {Zhao},
  \citenamefont {Zheng},\ and\ \citenamefont {Li}}]{yang2016programmable}%
  \BibitemOpen
  \bibfield  {author} {\bibinfo {author} {\bibfnamefont {H.}~\bibnamefont
  {Yang}}, \bibinfo {author} {\bibfnamefont {X.}~\bibnamefont {Cao}}, \bibinfo
  {author} {\bibfnamefont {F.}~\bibnamefont {Yang}}, \bibinfo {author}
  {\bibfnamefont {J.}~\bibnamefont {Gao}}, \bibinfo {author} {\bibfnamefont
  {S.}~\bibnamefont {Xu}}, \bibinfo {author} {\bibfnamefont {M.}~\bibnamefont
  {Li}}, \bibinfo {author} {\bibfnamefont {X.}~\bibnamefont {Chen}}, \bibinfo
  {author} {\bibfnamefont {Y.}~\bibnamefont {Zhao}}, \bibinfo {author}
  {\bibfnamefont {Y.}~\bibnamefont {Zheng}}, \ and\ \bibinfo {author}
  {\bibfnamefont {S.}~\bibnamefont {Li}},\ }\bibfield  {title} {\enquote
  {\bibinfo {title} {A programmable metasurface with dynamic polarization,
  scattering and focusing control},}\ }\href@noop {} {\bibfield  {journal}
  {\bibinfo  {journal} {Scientific reports}\ }\textbf {\bibinfo {volume} {6}},\
  \bibinfo {pages} {35692} (\bibinfo {year} {2016})}\BibitemShut {NoStop}%
\bibitem [{\citenamefont {Li}\ \emph {et~al.}(2016)\citenamefont {Li},
  \citenamefont {Li}, \citenamefont {Xu}, \citenamefont {Wu}, \citenamefont
  {Wu}, \citenamefont {Wan}, \citenamefont {Cheng},\ and\ \citenamefont
  {Cui}}]{li2016transmission}%
  \BibitemOpen
  \bibfield  {author} {\bibinfo {author} {\bibfnamefont {Y.~B.}\ \bibnamefont
  {Li}}, \bibinfo {author} {\bibfnamefont {L.~L.}\ \bibnamefont {Li}}, \bibinfo
  {author} {\bibfnamefont {B.~B.}\ \bibnamefont {Xu}}, \bibinfo {author}
  {\bibfnamefont {W.}~\bibnamefont {Wu}}, \bibinfo {author} {\bibfnamefont
  {R.~Y.}\ \bibnamefont {Wu}}, \bibinfo {author} {\bibfnamefont
  {X.}~\bibnamefont {Wan}}, \bibinfo {author} {\bibfnamefont {Q.}~\bibnamefont
  {Cheng}}, \ and\ \bibinfo {author} {\bibfnamefont {T.~J.}\ \bibnamefont
  {Cui}},\ }\bibfield  {title} {\enquote {\bibinfo {title} {Transmission-type
  2-bit programmable metasurface for single-sensor and single-frequency
  microwave imaging},}\ }\href@noop {} {\bibfield  {journal} {\bibinfo
  {journal} {Scientific reports}\ }\textbf {\bibinfo {volume} {6}},\ \bibinfo
  {pages} {23731} (\bibinfo {year} {2016})}\BibitemShut {NoStop}%
\bibitem [{\citenamefont {Xu}\ \emph {et~al.}(2016{\natexlab{b}})\citenamefont
  {Xu}, \citenamefont {Tang}, \citenamefont {Ma}, \citenamefont {Luo},
  \citenamefont {Cai}, \citenamefont {Sun}, \citenamefont {He},\ and\
  \citenamefont {Zhou}}]{xu2016tunable}%
  \BibitemOpen
  \bibfield  {author} {\bibinfo {author} {\bibfnamefont {H.-X.}\ \bibnamefont
  {Xu}}, \bibinfo {author} {\bibfnamefont {S.}~\bibnamefont {Tang}}, \bibinfo
  {author} {\bibfnamefont {S.}~\bibnamefont {Ma}}, \bibinfo {author}
  {\bibfnamefont {W.}~\bibnamefont {Luo}}, \bibinfo {author} {\bibfnamefont
  {T.}~\bibnamefont {Cai}}, \bibinfo {author} {\bibfnamefont {S.}~\bibnamefont
  {Sun}}, \bibinfo {author} {\bibfnamefont {Q.}~\bibnamefont {He}}, \ and\
  \bibinfo {author} {\bibfnamefont {L.}~\bibnamefont {Zhou}},\ }\bibfield
  {title} {\enquote {\bibinfo {title} {Tunable microwave metasurfaces for
  high-performance operations: dispersion compensation and dynamical switch},}\
  }\href@noop {} {\bibfield  {journal} {\bibinfo  {journal} {Scientific
  reports}\ }\textbf {\bibinfo {volume} {6}},\ \bibinfo {pages} {38255}
  (\bibinfo {year} {2016}{\natexlab{b}})}\BibitemShut {NoStop}%
\bibitem [{\citenamefont {del Hougne}\ \emph {et~al.}(2016)\citenamefont {del
  Hougne}, \citenamefont {Lemoult}, \citenamefont {Fink},\ and\ \citenamefont
  {Lerosey}}]{Hougne2016}%
  \BibitemOpen
  \bibfield  {author} {\bibinfo {author} {\bibfnamefont {P.}~\bibnamefont {del
  Hougne}}, \bibinfo {author} {\bibfnamefont {F.}~\bibnamefont {Lemoult}},
  \bibinfo {author} {\bibfnamefont {M.}~\bibnamefont {Fink}}, \ and\ \bibinfo
  {author} {\bibfnamefont {G.}~\bibnamefont {Lerosey}},\ }\bibfield  {title}
  {\enquote {\bibinfo {title} {Spatiotemporal wave front shaping in a microwave
  cavity},}\ }\href {\doibase 10.1103/PhysRevLett.117.134302} {\bibfield
  {journal} {\bibinfo  {journal} {Phys. Rev. Lett.}\ }\textbf {\bibinfo
  {volume} {117}},\ \bibinfo {pages} {134302} (\bibinfo {year}
  {2016})}\BibitemShut {NoStop}%
\bibitem [{\citenamefont {Xu}\ \emph {et~al.}(2016{\natexlab{c}})\citenamefont
  {Xu}, \citenamefont {Ma}, \citenamefont {Luo}, \citenamefont {Cai},
  \citenamefont {Sun}, \citenamefont {He},\ and\ \citenamefont
  {Zhou}}]{zhou2016focusing}%
  \BibitemOpen
  \bibfield  {author} {\bibinfo {author} {\bibfnamefont {H.-X.}\ \bibnamefont
  {Xu}}, \bibinfo {author} {\bibfnamefont {S.}~\bibnamefont {Ma}}, \bibinfo
  {author} {\bibfnamefont {W.}~\bibnamefont {Luo}}, \bibinfo {author}
  {\bibfnamefont {T.}~\bibnamefont {Cai}}, \bibinfo {author} {\bibfnamefont
  {S.}~\bibnamefont {Sun}}, \bibinfo {author} {\bibfnamefont {Q.}~\bibnamefont
  {He}}, \ and\ \bibinfo {author} {\bibfnamefont {L.}~\bibnamefont {Zhou}},\
  }\bibfield  {title} {\enquote {\bibinfo {title} {Aberration-free and
  functionality-switchable meta-lenses based on tunable metasurfaces},}\ }\href
  {\doibase 10.1063/1.4967438} {\bibfield  {journal} {\bibinfo  {journal}
  {Applied Physics Letters}\ }\textbf {\bibinfo {volume} {109}},\ \bibinfo
  {pages} {193506} (\bibinfo {year} {2016}{\natexlab{c}})},\ \Eprint
  {http://arxiv.org/abs/https://doi.org/10.1063/1.4967438}
  {https://doi.org/10.1063/1.4967438} \BibitemShut {NoStop}%
\bibitem [{\citenamefont {Chen}\ \emph {et~al.}(2017)\citenamefont {Chen},
  \citenamefont {Feng}, \citenamefont {Monticone}, \citenamefont {Zhao},
  \citenamefont {Zhu}, \citenamefont {Jiang}, \citenamefont {Zhang},
  \citenamefont {Kim}, \citenamefont {Ding}, \citenamefont {Zhang},
  \citenamefont {Al\`u},\ and\ \citenamefont {Qiu}}]{chen2017reconfigurable}%
  \BibitemOpen
  \bibfield  {author} {\bibinfo {author} {\bibfnamefont {K.}~\bibnamefont
  {Chen}}, \bibinfo {author} {\bibfnamefont {Y.}~\bibnamefont {Feng}}, \bibinfo
  {author} {\bibfnamefont {F.}~\bibnamefont {Monticone}}, \bibinfo {author}
  {\bibfnamefont {J.}~\bibnamefont {Zhao}}, \bibinfo {author} {\bibfnamefont
  {B.}~\bibnamefont {Zhu}}, \bibinfo {author} {\bibfnamefont {T.}~\bibnamefont
  {Jiang}}, \bibinfo {author} {\bibfnamefont {L.}~\bibnamefont {Zhang}},
  \bibinfo {author} {\bibfnamefont {Y.}~\bibnamefont {Kim}}, \bibinfo {author}
  {\bibfnamefont {X.}~\bibnamefont {Ding}}, \bibinfo {author} {\bibfnamefont
  {S.}~\bibnamefont {Zhang}}, \bibinfo {author} {\bibfnamefont
  {A.}~\bibnamefont {Al\`u}}, \ and\ \bibinfo {author} {\bibfnamefont {C.-W.}\
  \bibnamefont {Qiu}},\ }\bibfield  {title} {\enquote {\bibinfo {title} {A
  reconfigurable active huygens' metalens},}\ }\href {\doibase
  10.1002/adma.201606422} {\bibfield  {journal} {\bibinfo  {journal} {Advanced
  Materials}\ }\textbf {\bibinfo {volume} {29}},\ \bibinfo {pages} {1606422}
  (\bibinfo {year} {2017})}\BibitemShut {NoStop}%
\bibitem [{\citenamefont {Li}\ \emph {et~al.}(2017)\citenamefont {Li},
  \citenamefont {Cui}, \citenamefont {Ji}, \citenamefont {Liu}, \citenamefont
  {Ding}, \citenamefont {Wan}, \citenamefont {Li}, \citenamefont {Jiang},
  \citenamefont {Qiu},\ and\ \citenamefont {Zhang}}]{li2017electromagnetic}%
  \BibitemOpen
  \bibfield  {author} {\bibinfo {author} {\bibfnamefont {L.}~\bibnamefont
  {Li}}, \bibinfo {author} {\bibfnamefont {T.~J.}\ \bibnamefont {Cui}},
  \bibinfo {author} {\bibfnamefont {W.}~\bibnamefont {Ji}}, \bibinfo {author}
  {\bibfnamefont {S.}~\bibnamefont {Liu}}, \bibinfo {author} {\bibfnamefont
  {J.}~\bibnamefont {Ding}}, \bibinfo {author} {\bibfnamefont {X.}~\bibnamefont
  {Wan}}, \bibinfo {author} {\bibfnamefont {Y.~B.}\ \bibnamefont {Li}},
  \bibinfo {author} {\bibfnamefont {M.}~\bibnamefont {Jiang}}, \bibinfo
  {author} {\bibfnamefont {C.-W.}\ \bibnamefont {Qiu}}, \ and\ \bibinfo
  {author} {\bibfnamefont {S.}~\bibnamefont {Zhang}},\ }\bibfield  {title}
  {\enquote {\bibinfo {title} {Electromagnetic reprogrammable
  coding-metasurface holograms},}\ }\href@noop {} {\bibfield  {journal}
  {\bibinfo  {journal} {Nature communications}\ }\textbf {\bibinfo {volume}
  {8}},\ \bibinfo {pages} {1--7} (\bibinfo {year} {2017})}\BibitemShut
  {NoStop}%
\bibitem [{\citenamefont {Ratni}\ \emph {et~al.}(2018)\citenamefont {Ratni},
  \citenamefont {de~Lustrac}, \citenamefont {Piau},\ and\ \citenamefont
  {Burokur}}]{Ratni18}%
  \BibitemOpen
  \bibfield  {author} {\bibinfo {author} {\bibfnamefont {B.}~\bibnamefont
  {Ratni}}, \bibinfo {author} {\bibfnamefont {A.}~\bibnamefont {de~Lustrac}},
  \bibinfo {author} {\bibfnamefont {G.-P.}\ \bibnamefont {Piau}}, \ and\
  \bibinfo {author} {\bibfnamefont {S.~N.}\ \bibnamefont {Burokur}},\
  }\bibfield  {title} {\enquote {\bibinfo {title} {Reconfigurable meta-mirror
  for wavefronts control: applications to microwave antennas},}\ }\href
  {\doibase 10.1364/OE.26.002613} {\bibfield  {journal} {\bibinfo  {journal}
  {Opt. Express}\ }\textbf {\bibinfo {volume} {26}},\ \bibinfo {pages}
  {2613--2624} (\bibinfo {year} {2018})}\BibitemShut {NoStop}%
\bibitem [{\citenamefont {del Hougne}\ \emph {et~al.}(2018)\citenamefont {del
  Hougne}, \citenamefont {Imani}, \citenamefont {Sleasman}, \citenamefont
  {Gollub}, \citenamefont {Fink}, \citenamefont {Lerosey},\ and\ \citenamefont
  {Smith}}]{Hougne2018dynamic}%
  \BibitemOpen
  \bibfield  {author} {\bibinfo {author} {\bibfnamefont {P.}~\bibnamefont {del
  Hougne}}, \bibinfo {author} {\bibfnamefont {M.~F}\ \bibnamefont {Imani}},
  \bibinfo {author} {\bibfnamefont {T.}~\bibnamefont {Sleasman}}, \bibinfo
  {author} {\bibfnamefont {J.~N.}\ \bibnamefont {Gollub}}, \bibinfo {author}
  {\bibfnamefont {M.}~\bibnamefont {Fink}}, \bibinfo {author} {\bibfnamefont
  {G.}~\bibnamefont {Lerosey}}, \ and\ \bibinfo {author} {\bibfnamefont
  {D.~R.}\ \bibnamefont {Smith}},\ }\bibfield  {title} {\enquote {\bibinfo
  {title} {Dynamic metasurface aperture as smart around-the-corner motion
  detector},}\ }\href@noop {} {\bibfield  {journal} {\bibinfo  {journal}
  {Scientific reports}\ }\textbf {\bibinfo {volume} {8}},\ \bibinfo {pages}
  {1--10} (\bibinfo {year} {2018})}\BibitemShut {NoStop}%
\bibitem [{\citenamefont {Komar}\ \emph {et~al.}(2018)\citenamefont {Komar},
  \citenamefont {Paniagua-Dominguez}, \citenamefont {Miroshnichenko},
  \citenamefont {Yu}, \citenamefont {Kivshar}, \citenamefont {Kuznetsov},\ and\
  \citenamefont {Neshev}}]{komar2018dynamic}%
  \BibitemOpen
  \bibfield  {author} {\bibinfo {author} {\bibfnamefont {A.}~\bibnamefont
  {Komar}}, \bibinfo {author} {\bibfnamefont {R.}~\bibnamefont
  {Paniagua-Dominguez}}, \bibinfo {author} {\bibfnamefont {A.}~\bibnamefont
  {Miroshnichenko}}, \bibinfo {author} {\bibfnamefont {Y.~F.}\ \bibnamefont
  {Yu}}, \bibinfo {author} {\bibfnamefont {Y.~S.}\ \bibnamefont {Kivshar}},
  \bibinfo {author} {\bibfnamefont {A.~I.}\ \bibnamefont {Kuznetsov}}, \ and\
  \bibinfo {author} {\bibfnamefont {D.}~\bibnamefont {Neshev}},\ }\bibfield
  {title} {\enquote {\bibinfo {title} {Dynamic beam switching by liquid crystal
  tunable dielectric metasurfaces},}\ }\href@noop {} {\bibfield  {journal}
  {\bibinfo  {journal} {ACS Photonics}\ }\textbf {\bibinfo {volume} {5}},\
  \bibinfo {pages} {1742--1748} (\bibinfo {year} {2018})}\BibitemShut {NoStop}%
\bibitem [{\citenamefont {del Hougne}\ and\ \citenamefont
  {Lerosey}(2018)}]{Hougne2018}%
  \BibitemOpen
  \bibfield  {author} {\bibinfo {author} {\bibfnamefont {P.}~\bibnamefont {del
  Hougne}}\ and\ \bibinfo {author} {\bibfnamefont {G.}~\bibnamefont
  {Lerosey}},\ }\bibfield  {title} {\enquote {\bibinfo {title} {Leveraging
  chaos for wave-based analog computation: Demonstration with indoor wireless
  communication signals},}\ }\href {\doibase 10.1103/PhysRevX.8.041037}
  {\bibfield  {journal} {\bibinfo  {journal} {Phys. Rev. X}\ }\textbf {\bibinfo
  {volume} {8}},\ \bibinfo {pages} {041037} (\bibinfo {year}
  {2018})}\BibitemShut {NoStop}%
\bibitem [{\citenamefont {Li}\ \emph {et~al.}(2019)\citenamefont {Li},
  \citenamefont {Shuang}, \citenamefont {Ma}, \citenamefont {Li}, \citenamefont
  {Zhao}, \citenamefont {Wei}, \citenamefont {Liu}, \citenamefont {Hao},
  \citenamefont {Qiu},\ and\ \citenamefont {Cui}}]{li2019intelligent}%
  \BibitemOpen
  \bibfield  {author} {\bibinfo {author} {\bibfnamefont {L.}~\bibnamefont
  {Li}}, \bibinfo {author} {\bibfnamefont {Y.}~\bibnamefont {Shuang}}, \bibinfo
  {author} {\bibfnamefont {Q.}~\bibnamefont {Ma}}, \bibinfo {author}
  {\bibfnamefont {H.}~\bibnamefont {Li}}, \bibinfo {author} {\bibfnamefont
  {H.}~\bibnamefont {Zhao}}, \bibinfo {author} {\bibfnamefont {M.}~\bibnamefont
  {Wei}}, \bibinfo {author} {\bibfnamefont {C.}~\bibnamefont {Liu}}, \bibinfo
  {author} {\bibfnamefont {C.}~\bibnamefont {Hao}}, \bibinfo {author}
  {\bibfnamefont {C.-W.}\ \bibnamefont {Qiu}}, \ and\ \bibinfo {author}
  {\bibfnamefont {T.~J.}\ \bibnamefont {Cui}},\ }\bibfield  {title} {\enquote
  {\bibinfo {title} {Intelligent metasurface imager and recognizer},}\
  }\href@noop {} {\bibfield  {journal} {\bibinfo  {journal} {Light: Science \&
  Applications}\ }\textbf {\bibinfo {volume} {8}},\ \bibinfo {pages} {1--9}
  (\bibinfo {year} {2019})}\BibitemShut {NoStop}%
\bibitem [{\citenamefont {Ma}\ \emph {et~al.}(2019)\citenamefont {Ma},
  \citenamefont {Bai}, \citenamefont {Jing}, \citenamefont {Yang},
  \citenamefont {Li},\ and\ \citenamefont {Cui}}]{ma2019smart}%
  \BibitemOpen
  \bibfield  {author} {\bibinfo {author} {\bibfnamefont {Q.}~\bibnamefont
  {Ma}}, \bibinfo {author} {\bibfnamefont {G.~D.}\ \bibnamefont {Bai}},
  \bibinfo {author} {\bibfnamefont {H.~B.}\ \bibnamefont {Jing}}, \bibinfo
  {author} {\bibfnamefont {C.}~\bibnamefont {Yang}}, \bibinfo {author}
  {\bibfnamefont {L.}~\bibnamefont {Li}}, \ and\ \bibinfo {author}
  {\bibfnamefont {T.~J.}\ \bibnamefont {Cui}},\ }\bibfield  {title} {\enquote
  {\bibinfo {title} {Smart metasurface with self-adaptively reprogrammable
  functions},}\ }\href@noop {} {\bibfield  {journal} {\bibinfo  {journal}
  {Light: Science \& Applications}\ }\textbf {\bibinfo {volume} {8}},\ \bibinfo
  {pages} {1--12} (\bibinfo {year} {2019})}\BibitemShut {NoStop}%
\bibitem [{\citenamefont {Liu}\ \emph {et~al.}(2019)\citenamefont {Liu},
  \citenamefont {Tsilipakos}, \citenamefont {Pitilakis}, \citenamefont
  {Tasolamprou}, \citenamefont {Mirmoosa}, \citenamefont {Kantartzis},
  \citenamefont {Kwon}, \citenamefont {Georgiou}, \citenamefont {Kossifos},
  \citenamefont {Antoniades}, \citenamefont {Kafesaki}, \citenamefont
  {Soukoulis},\ and\ \citenamefont {Tretyakov}}]{Tretyakov_IntMS2019}%
  \BibitemOpen
  \bibfield  {author} {\bibinfo {author} {\bibfnamefont {F.}~\bibnamefont
  {Liu}}, \bibinfo {author} {\bibfnamefont {O.}~\bibnamefont {Tsilipakos}},
  \bibinfo {author} {\bibfnamefont {A.}~\bibnamefont {Pitilakis}}, \bibinfo
  {author} {\bibfnamefont {A.~C.}\ \bibnamefont {Tasolamprou}}, \bibinfo
  {author} {\bibfnamefont {M.~S.}\ \bibnamefont {Mirmoosa}}, \bibinfo {author}
  {\bibfnamefont {N.~V.}\ \bibnamefont {Kantartzis}}, \bibinfo {author}
  {\bibfnamefont {D-H.}\ \bibnamefont {Kwon}}, \bibinfo {author} {\bibfnamefont
  {J.}~\bibnamefont {Georgiou}}, \bibinfo {author} {\bibfnamefont
  {K.}~\bibnamefont {Kossifos}}, \bibinfo {author} {\bibfnamefont {M.~A.}\
  \bibnamefont {Antoniades}}, \bibinfo {author} {\bibfnamefont
  {M.}~\bibnamefont {Kafesaki}}, \bibinfo {author} {\bibfnamefont {C.~M.}\
  \bibnamefont {Soukoulis}}, \ and\ \bibinfo {author} {\bibfnamefont {S.~A.}\
  \bibnamefont {Tretyakov}},\ }\bibfield  {title} {\enquote {\bibinfo {title}
  {Intelligent metasurfaces with continuously tunable local surface impedance
  for multiple reconfigurable functions},}\ }\href {\doibase
  10.1103/PhysRevApplied.11.044024} {\bibfield  {journal} {\bibinfo  {journal}
  {Phys. Rev. Applied}\ }\textbf {\bibinfo {volume} {11}},\ \bibinfo {pages}
  {044024} (\bibinfo {year} {2019})}\BibitemShut {NoStop}%
\bibitem [{\citenamefont {del Hougne}\ \emph {et~al.}(2019)\citenamefont {del
  Hougne}, \citenamefont {Fink},\ and\ \citenamefont {Lerosey}}]{Hougne2019}%
  \BibitemOpen
  \bibfield  {author} {\bibinfo {author} {\bibfnamefont {P.}~\bibnamefont {del
  Hougne}}, \bibinfo {author} {\bibfnamefont {M.}~\bibnamefont {Fink}}, \ and\
  \bibinfo {author} {\bibfnamefont {G.}~\bibnamefont {Lerosey}},\ }\bibfield
  {title} {\enquote {\bibinfo {title} {Optimally diverse communication channels
  in disordered environments with tuned randomness},}\ }\href@noop {}
  {\bibfield  {journal} {\bibinfo  {journal} {Nature Electronics}\ }\textbf
  {\bibinfo {volume} {2}},\ \bibinfo {pages} {36--41} (\bibinfo {year}
  {2019})}\BibitemShut {NoStop}%
\bibitem [{\citenamefont {Leitis}\ \emph {et~al.}(2020)\citenamefont {Leitis},
  \citenamefont {Heßler}, \citenamefont {Wahl}, \citenamefont {Wuttig},
  \citenamefont {Taubner}, \citenamefont {Tittl},\ and\ \citenamefont
  {Altug}}]{Altug2020}%
  \BibitemOpen
  \bibfield  {author} {\bibinfo {author} {\bibfnamefont {A.}~\bibnamefont
  {Leitis}}, \bibinfo {author} {\bibfnamefont {A.}~\bibnamefont {Heßler}},
  \bibinfo {author} {\bibfnamefont {S.}~\bibnamefont {Wahl}}, \bibinfo {author}
  {\bibfnamefont {M.}~\bibnamefont {Wuttig}}, \bibinfo {author} {\bibfnamefont
  {T.}~\bibnamefont {Taubner}}, \bibinfo {author} {\bibfnamefont
  {A.}~\bibnamefont {Tittl}}, \ and\ \bibinfo {author} {\bibfnamefont
  {H.}~\bibnamefont {Altug}},\ }\bibfield  {title} {\enquote {\bibinfo {title}
  {All-dielectric programmable huygens' metasurfaces},}\ }\href {\doibase
  10.1002/adfm.201910259} {\bibfield  {journal} {\bibinfo  {journal} {Advanced
  Functional Materials}\ ,\ \bibinfo {pages} {1910259}} (\bibinfo {year}
  {2020})}\BibitemShut {NoStop}%
\bibitem [{\citenamefont {Ratni}\ \emph {et~al.}(2020)\citenamefont {Ratni},
  \citenamefont {Wang}, \citenamefont {Zhang}, \citenamefont {Ding},
  \citenamefont {de~Lustrac}, \citenamefont {Piau},\ and\ \citenamefont
  {Burokur}}]{Ratni2020}%
  \BibitemOpen
  \bibfield  {author} {\bibinfo {author} {\bibfnamefont {B.}~\bibnamefont
  {Ratni}}, \bibinfo {author} {\bibfnamefont {Z.}~\bibnamefont {Wang}},
  \bibinfo {author} {\bibfnamefont {K.}~\bibnamefont {Zhang}}, \bibinfo
  {author} {\bibfnamefont {X.}~\bibnamefont {Ding}}, \bibinfo {author}
  {\bibfnamefont {A.}~\bibnamefont {de~Lustrac}}, \bibinfo {author}
  {\bibfnamefont {G.-P.}\ \bibnamefont {Piau}}, \ and\ \bibinfo {author}
  {\bibfnamefont {S.~N.}\ \bibnamefont {Burokur}},\ }\bibfield  {title}
  {\enquote {\bibinfo {title} {Dynamically controlling spatial energy
  distribution with a holographic metamirror for adaptive focusing},}\ }\href
  {\doibase 10.1103/PhysRevApplied.13.034006} {\bibfield  {journal} {\bibinfo
  {journal} {Phys. Rev. Applied}\ }\textbf {\bibinfo {volume} {13}},\ \bibinfo
  {pages} {034006} (\bibinfo {year} {2020})}\BibitemShut {NoStop}%
\bibitem [{\citenamefont {Hadad}\ \emph {et~al.}(2015)\citenamefont {Hadad},
  \citenamefont {Sounas},\ and\ \citenamefont
  {Al\`u}}]{Alu2015_time_modulated}%
  \BibitemOpen
  \bibfield  {author} {\bibinfo {author} {\bibfnamefont {Y.}~\bibnamefont
  {Hadad}}, \bibinfo {author} {\bibfnamefont {D.~L.}\ \bibnamefont {Sounas}}, \
  and\ \bibinfo {author} {\bibfnamefont {A.}~\bibnamefont {Al\`u}},\ }\bibfield
   {title} {\enquote {\bibinfo {title} {Space-time gradient metasurfaces},}\
  }\href {\doibase 10.1103/PhysRevB.92.100304} {\bibfield  {journal} {\bibinfo
  {journal} {Phys. Rev. B}\ }\textbf {\bibinfo {volume} {92}},\ \bibinfo
  {pages} {100304} (\bibinfo {year} {2015})}\BibitemShut {NoStop}%
\bibitem [{\citenamefont {Zhao}\ \emph {et~al.}(2018)\citenamefont {Zhao},
  \citenamefont {Yang}, \citenamefont {Dai}, \citenamefont {Cheng},
  \citenamefont {Li}, \citenamefont {Qi}, \citenamefont {Ke}, \citenamefont
  {Bai}, \citenamefont {Liu}, \citenamefont {Jin}, \citenamefont {Al\`u},\ and\
  \citenamefont {Cui}}]{Alu2018_time_modulated}%
  \BibitemOpen
  \bibfield  {author} {\bibinfo {author} {\bibfnamefont {J.}~\bibnamefont
  {Zhao}}, \bibinfo {author} {\bibfnamefont {X.}~\bibnamefont {Yang}}, \bibinfo
  {author} {\bibfnamefont {J.~Y.}\ \bibnamefont {Dai}}, \bibinfo {author}
  {\bibfnamefont {Q.}~\bibnamefont {Cheng}}, \bibinfo {author} {\bibfnamefont
  {X.}~\bibnamefont {Li}}, \bibinfo {author} {\bibfnamefont {N.~H.}\
  \bibnamefont {Qi}}, \bibinfo {author} {\bibfnamefont {J.~C.}\ \bibnamefont
  {Ke}}, \bibinfo {author} {\bibfnamefont {G.~D.}\ \bibnamefont {Bai}},
  \bibinfo {author} {\bibfnamefont {S.}~\bibnamefont {Liu}}, \bibinfo {author}
  {\bibfnamefont {S.}~\bibnamefont {Jin}}, \bibinfo {author} {\bibfnamefont
  {A.}~\bibnamefont {Al\`u}}, \ and\ \bibinfo {author} {\bibfnamefont {T.~J.}\
  \bibnamefont {Cui}},\ }\bibfield  {title} {\enquote {\bibinfo {title}
  {Programmable time-domain digital-coding metasurface for non-linear harmonic
  manipulation and new wireless communication systems},}\ }\href {\doibase
  10.1093/nsr/nwy135} {\bibfield  {journal} {\bibinfo  {journal} {National
  Science Review}\ }\textbf {\bibinfo {volume} {6}},\ \bibinfo {pages}
  {231--238} (\bibinfo {year} {2018})}\BibitemShut {NoStop}%
\bibitem [{\citenamefont {Zhang}\ \emph {et~al.}(2018)\citenamefont {Zhang},
  \citenamefont {Chen}, \citenamefont {Liu}, \citenamefont {Zhang},
  \citenamefont {Zhao}, \citenamefont {Dai}, \citenamefont {Bai}, \citenamefont
  {Wan}, \citenamefont {Cheng}, \citenamefont {Castaldi}, \citenamefont
  {Galdi},\ and\ \citenamefont {Cui}}]{Zhang2018_time_modulated}%
  \BibitemOpen
  \bibfield  {author} {\bibinfo {author} {\bibfnamefont {L.}~\bibnamefont
  {Zhang}}, \bibinfo {author} {\bibfnamefont {X.~Q.}\ \bibnamefont {Chen}},
  \bibinfo {author} {\bibfnamefont {S.}~\bibnamefont {Liu}}, \bibinfo {author}
  {\bibfnamefont {Q.}~\bibnamefont {Zhang}}, \bibinfo {author} {\bibfnamefont
  {J.}~\bibnamefont {Zhao}}, \bibinfo {author} {\bibfnamefont {J.~Y.}\
  \bibnamefont {Dai}}, \bibinfo {author} {\bibfnamefont {G.~D.}\ \bibnamefont
  {Bai}}, \bibinfo {author} {\bibfnamefont {X.}~\bibnamefont {Wan}}, \bibinfo
  {author} {\bibfnamefont {Q.}~\bibnamefont {Cheng}}, \bibinfo {author}
  {\bibfnamefont {G.}~\bibnamefont {Castaldi}}, \bibinfo {author}
  {\bibfnamefont {V.}~\bibnamefont {Galdi}}, \ and\ \bibinfo {author}
  {\bibfnamefont {T.~J.}\ \bibnamefont {Cui}},\ }\bibfield  {title} {\enquote
  {\bibinfo {title} {Space-time-coding digital metasurfaces},}\ }\href@noop {}
  {\bibfield  {journal} {\bibinfo  {journal} {Nature communications}\ }\textbf
  {\bibinfo {volume} {9}},\ \bibinfo {pages} {1--11} (\bibinfo {year}
  {2018})}\BibitemShut {NoStop}%
\bibitem [{\citenamefont {Liu}\ \emph {et~al.}(2018)\citenamefont {Liu},
  \citenamefont {Powell}, \citenamefont {Zarate},\ and\ \citenamefont
  {Shadrivov}}]{Shadrivov2018_time_modulated}%
  \BibitemOpen
  \bibfield  {author} {\bibinfo {author} {\bibfnamefont {M.}~\bibnamefont
  {Liu}}, \bibinfo {author} {\bibfnamefont {D.~A.}\ \bibnamefont {Powell}},
  \bibinfo {author} {\bibfnamefont {Y.}~\bibnamefont {Zarate}}, \ and\ \bibinfo
  {author} {\bibfnamefont {I.~V.}\ \bibnamefont {Shadrivov}},\ }\bibfield
  {title} {\enquote {\bibinfo {title} {Huygens' metadevices for parametric
  waves},}\ }\href {\doibase 10.1103/PhysRevX.8.031077} {\bibfield  {journal}
  {\bibinfo  {journal} {Phys. Rev. X}\ }\textbf {\bibinfo {volume} {8}},\
  \bibinfo {pages} {031077} (\bibinfo {year} {2018})}\BibitemShut {NoStop}%
\bibitem [{\citenamefont {Salary}\ \emph {et~al.}(2018)\citenamefont {Salary},
  \citenamefont {Jafar-Zanjani},\ and\ \citenamefont
  {Mosallaei}}]{Salary2018_Nonreciprocity_time_modulated}%
  \BibitemOpen
  \bibfield  {author} {\bibinfo {author} {\bibfnamefont {M.~M.}\ \bibnamefont
  {Salary}}, \bibinfo {author} {\bibfnamefont {S.}~\bibnamefont
  {Jafar-Zanjani}}, \ and\ \bibinfo {author} {\bibfnamefont {H.}~\bibnamefont
  {Mosallaei}},\ }\bibfield  {title} {\enquote {\bibinfo {title} {Electrically
  tunable harmonics in time-modulated metasurfaces for wavefront
  engineering},}\ }\href {\doibase 10.1088/1367-2630/aaf47a} {\bibfield
  {journal} {\bibinfo  {journal} {New Journal of Physics}\ }\textbf {\bibinfo
  {volume} {20}},\ \bibinfo {pages} {123023} (\bibinfo {year}
  {2018})}\BibitemShut {NoStop}%
\bibitem [{\citenamefont {Caloz}\ \emph {et~al.}(2018)\citenamefont {Caloz},
  \citenamefont {Al\`u}, \citenamefont {Tretyakov}, \citenamefont {Sounas},
  \citenamefont {Achouri},\ and\ \citenamefont
  {Deck-L\'eger}}]{Tretyakov2018_Nonreciprocity}%
  \BibitemOpen
  \bibfield  {author} {\bibinfo {author} {\bibfnamefont {C.}~\bibnamefont
  {Caloz}}, \bibinfo {author} {\bibfnamefont {A.}~\bibnamefont {Al\`u}},
  \bibinfo {author} {\bibfnamefont {S.}~\bibnamefont {Tretyakov}}, \bibinfo
  {author} {\bibfnamefont {D.}~\bibnamefont {Sounas}}, \bibinfo {author}
  {\bibfnamefont {K.}~\bibnamefont {Achouri}}, \ and\ \bibinfo {author}
  {\bibfnamefont {Z.-L.}\ \bibnamefont {Deck-L\'eger}},\ }\bibfield  {title}
  {\enquote {\bibinfo {title} {Electromagnetic nonreciprocity},}\ }\href
  {\doibase 10.1103/PhysRevApplied.10.047001} {\bibfield  {journal} {\bibinfo
  {journal} {Phys. Rev. Applied}\ }\textbf {\bibinfo {volume} {10}},\ \bibinfo
  {pages} {047001} (\bibinfo {year} {2018})}\BibitemShut {NoStop}%
\bibitem [{\citenamefont {Zang}\ \emph
  {et~al.}(2019{\natexlab{a}})\citenamefont {Zang}, \citenamefont
  {Correas-Serrano}, \citenamefont {Do}, \citenamefont {Liu}, \citenamefont
  {Alvarez-Melcon},\ and\ \citenamefont
  {Gomez-Diaz}}]{Gomez-Diaz2019_Nonreciprocity_time_modulated}%
  \BibitemOpen
  \bibfield  {author} {\bibinfo {author} {\bibfnamefont {J.W.}\ \bibnamefont
  {Zang}}, \bibinfo {author} {\bibfnamefont {D.}~\bibnamefont
  {Correas-Serrano}}, \bibinfo {author} {\bibfnamefont {J.T.S.}\ \bibnamefont
  {Do}}, \bibinfo {author} {\bibfnamefont {X.}~\bibnamefont {Liu}}, \bibinfo
  {author} {\bibfnamefont {A.}~\bibnamefont {Alvarez-Melcon}}, \ and\ \bibinfo
  {author} {\bibfnamefont {J.S.}\ \bibnamefont {Gomez-Diaz}},\ }\bibfield
  {title} {\enquote {\bibinfo {title} {Nonreciprocal wavefront engineering with
  time-modulated gradient metasurfaces},}\ }\href {\doibase
  10.1103/PhysRevApplied.11.054054} {\bibfield  {journal} {\bibinfo  {journal}
  {Phys. Rev. Applied}\ }\textbf {\bibinfo {volume} {11}},\ \bibinfo {pages}
  {054054} (\bibinfo {year} {2019}{\natexlab{a}})}\BibitemShut {NoStop}%
\bibitem [{\citenamefont {Zang}\ \emph
  {et~al.}(2019{\natexlab{b}})\citenamefont {Zang}, \citenamefont
  {Alvarez-Melcon},\ and\ \citenamefont
  {Gomez-Diaz}}]{Gomez-Diaz2019_Nonreciprocity}%
  \BibitemOpen
  \bibfield  {author} {\bibinfo {author} {\bibfnamefont {J.W.}\ \bibnamefont
  {Zang}}, \bibinfo {author} {\bibfnamefont {A.}~\bibnamefont
  {Alvarez-Melcon}}, \ and\ \bibinfo {author} {\bibfnamefont {J.S.}\
  \bibnamefont {Gomez-Diaz}},\ }\bibfield  {title} {\enquote {\bibinfo {title}
  {Nonreciprocal phased-array antennas},}\ }\href {\doibase
  10.1103/PhysRevApplied.12.054008} {\bibfield  {journal} {\bibinfo  {journal}
  {Phys. Rev. Applied}\ }\textbf {\bibinfo {volume} {12}},\ \bibinfo {pages}
  {054008} (\bibinfo {year} {2019}{\natexlab{b}})}\BibitemShut {NoStop}%
\bibitem [{\citenamefont {Mirmoosa}\ \emph {et~al.}(2019)\citenamefont
  {Mirmoosa}, \citenamefont {Ptitcyn}, \citenamefont {Asadchy},\ and\
  \citenamefont {Tretyakov}}]{Asadchy2019_time_modulated}%
  \BibitemOpen
  \bibfield  {author} {\bibinfo {author} {\bibfnamefont {M.S.}\ \bibnamefont
  {Mirmoosa}}, \bibinfo {author} {\bibfnamefont {G.A.}\ \bibnamefont
  {Ptitcyn}}, \bibinfo {author} {\bibfnamefont {V.S.}\ \bibnamefont {Asadchy}},
  \ and\ \bibinfo {author} {\bibfnamefont {S.A.}\ \bibnamefont {Tretyakov}},\
  }\bibfield  {title} {\enquote {\bibinfo {title} {Time-varying reactive
  elements for extreme accumulation of electromagnetic energy},}\ }\href
  {\doibase 10.1103/PhysRevApplied.11.014024} {\bibfield  {journal} {\bibinfo
  {journal} {Phys. Rev. Applied}\ }\textbf {\bibinfo {volume} {11}},\ \bibinfo
  {pages} {014024} (\bibinfo {year} {2019})}\BibitemShut {NoStop}%
\bibitem [{\citenamefont {Ptitcyn}\ \emph {et~al.}(2019)\citenamefont
  {Ptitcyn}, \citenamefont {Mirmoosa},\ and\ \citenamefont
  {Tretyakov}}]{Ptitcyn2019_time_modulated}%
  \BibitemOpen
  \bibfield  {author} {\bibinfo {author} {\bibfnamefont {G.}~\bibnamefont
  {Ptitcyn}}, \bibinfo {author} {\bibfnamefont {M.~S.}\ \bibnamefont
  {Mirmoosa}}, \ and\ \bibinfo {author} {\bibfnamefont {S.~A.}\ \bibnamefont
  {Tretyakov}},\ }\bibfield  {title} {\enquote {\bibinfo {title}
  {Time-modulated meta-atoms},}\ }\href {\doibase
  10.1103/PhysRevResearch.1.023014} {\bibfield  {journal} {\bibinfo  {journal}
  {Phys. Rev. Research}\ }\textbf {\bibinfo {volume} {1}},\ \bibinfo {pages}
  {023014} (\bibinfo {year} {2019})}\BibitemShut {NoStop}%
\bibitem [{\citenamefont {Wang}\ \emph {et~al.}(2016)\citenamefont {Wang},
  \citenamefont {Rogers}, \citenamefont {Gholipour}, \citenamefont {Wang},
  \citenamefont {Yuan}, \citenamefont {Teng},\ and\ \citenamefont
  {Zheludev}}]{wang2016optically}%
  \BibitemOpen
  \bibfield  {author} {\bibinfo {author} {\bibfnamefont {Q.}~\bibnamefont
  {Wang}}, \bibinfo {author} {\bibfnamefont {E.~T.F.}\ \bibnamefont {Rogers}},
  \bibinfo {author} {\bibfnamefont {B.}~\bibnamefont {Gholipour}}, \bibinfo
  {author} {\bibfnamefont {C-M.}\ \bibnamefont {Wang}}, \bibinfo {author}
  {\bibfnamefont {G.}~\bibnamefont {Yuan}}, \bibinfo {author} {\bibfnamefont
  {J.}~\bibnamefont {Teng}}, \ and\ \bibinfo {author} {\bibfnamefont {N.~I.}\
  \bibnamefont {Zheludev}},\ }\bibfield  {title} {\enquote {\bibinfo {title}
  {Optically reconfigurable metasurfaces and photonic devices based on phase
  change materials},}\ }\href@noop {} {\bibfield  {journal} {\bibinfo
  {journal} {Nature Photonics}\ }\textbf {\bibinfo {volume} {10}},\ \bibinfo
  {pages} {60} (\bibinfo {year} {2016})}\BibitemShut {NoStop}%
\bibitem [{\citenamefont {Huang}\ \emph {et~al.}(2016)\citenamefont {Huang},
  \citenamefont {Lee}, \citenamefont {Sokhoyan}, \citenamefont {Pala},
  \citenamefont {Thyagarajan}, \citenamefont {Han}, \citenamefont {Tsai},\ and\
  \citenamefont {Atwater}}]{huang2016gate}%
  \BibitemOpen
  \bibfield  {author} {\bibinfo {author} {\bibfnamefont {Y.-W.}\ \bibnamefont
  {Huang}}, \bibinfo {author} {\bibfnamefont {H.~W.~H.}\ \bibnamefont {Lee}},
  \bibinfo {author} {\bibfnamefont {R.}~\bibnamefont {Sokhoyan}}, \bibinfo
  {author} {\bibfnamefont {R.~A.}\ \bibnamefont {Pala}}, \bibinfo {author}
  {\bibfnamefont {K.}~\bibnamefont {Thyagarajan}}, \bibinfo {author}
  {\bibfnamefont {S.}~\bibnamefont {Han}}, \bibinfo {author} {\bibfnamefont
  {D.~P.}\ \bibnamefont {Tsai}}, \ and\ \bibinfo {author} {\bibfnamefont
  {H.~A.}\ \bibnamefont {Atwater}},\ }\bibfield  {title} {\enquote {\bibinfo
  {title} {Gate-tunable conducting oxide metasurfaces},}\ }\href@noop {}
  {\bibfield  {journal} {\bibinfo  {journal} {Nano letters}\ }\textbf {\bibinfo
  {volume} {16}},\ \bibinfo {pages} {5319--5325} (\bibinfo {year}
  {2016})}\BibitemShut {NoStop}%
\bibitem [{\citenamefont {Cui}\ \emph {et~al.}(2019)\citenamefont {Cui},
  \citenamefont {Bai},\ and\ \citenamefont {Sun}}]{Cui2019_Review_optics}%
  \BibitemOpen
  \bibfield  {author} {\bibinfo {author} {\bibfnamefont {T.}~\bibnamefont
  {Cui}}, \bibinfo {author} {\bibfnamefont {B.}~\bibnamefont {Bai}}, \ and\
  \bibinfo {author} {\bibfnamefont {H.-B.}\ \bibnamefont {Sun}},\ }\bibfield
  {title} {\enquote {\bibinfo {title} {Tunable metasurfaces based on active
  materials},}\ }\href {\doibase 10.1002/adfm.201806692} {\bibfield  {journal}
  {\bibinfo  {journal} {Advanced Functional Materials}\ }\textbf {\bibinfo
  {volume} {29}},\ \bibinfo {pages} {1806692} (\bibinfo {year}
  {2019})}\BibitemShut {NoStop}%
\bibitem [{\citenamefont {He}\ \emph {et~al.}(2019)\citenamefont {He},
  \citenamefont {Sun},\ and\ \citenamefont {Zhou}}]{he2019tunable}%
  \BibitemOpen
  \bibfield  {author} {\bibinfo {author} {\bibfnamefont {Q.}~\bibnamefont
  {He}}, \bibinfo {author} {\bibfnamefont {S.}~\bibnamefont {Sun}}, \ and\
  \bibinfo {author} {\bibfnamefont {L.}~\bibnamefont {Zhou}},\ }\bibfield
  {title} {\enquote {\bibinfo {title} {Tunable/reconfigurable metasurfaces:
  Physics and applications},}\ }\href@noop {} {\bibfield  {journal} {\bibinfo
  {journal} {Research}\ }\textbf {\bibinfo {volume} {2019}},\ \bibinfo {pages}
  {1849272} (\bibinfo {year} {2019})}\BibitemShut {NoStop}%
\bibitem [{\citenamefont {Pfeiffer}\ and\ \citenamefont
  {Grbic}(2013)}]{Grbic2013}%
  \BibitemOpen
  \bibfield  {author} {\bibinfo {author} {\bibfnamefont {C.}~\bibnamefont
  {Pfeiffer}}\ and\ \bibinfo {author} {\bibfnamefont {A.}~\bibnamefont
  {Grbic}},\ }\bibfield  {title} {\enquote {\bibinfo {title} {Metamaterial
  huygens' surfaces: Tailoring wave fronts with reflectionless sheets},}\
  }\href {\doibase 10.1103/PhysRevLett.110.197401} {\bibfield  {journal}
  {\bibinfo  {journal} {Phys. Rev. Lett.}\ }\textbf {\bibinfo {volume} {110}},\
  \bibinfo {pages} {197401} (\bibinfo {year} {2013})}\BibitemShut {NoStop}%
\bibitem [{\citenamefont {Epstein}\ and\ \citenamefont
  {Eleftheriades}(2014)}]{Epstein2014_ieee}%
  \BibitemOpen
  \bibfield  {author} {\bibinfo {author} {\bibfnamefont {A.}~\bibnamefont
  {Epstein}}\ and\ \bibinfo {author} {\bibfnamefont {G.~V.}\ \bibnamefont
  {Eleftheriades}},\ }\bibfield  {title} {\enquote {\bibinfo {title} {Passive
  lossless huygens metasurfaces for conversion of arbitrary source field to
  directive radiation},}\ }\href {\doibase 10.1109/TAP.2014.2354419} {\bibfield
   {journal} {\bibinfo  {journal} {IEEE Transactions on Antennas and
  Propagation}\ }\textbf {\bibinfo {volume} {62}},\ \bibinfo {pages}
  {5680--5695} (\bibinfo {year} {2014})}\BibitemShut {NoStop}%
\bibitem [{\citenamefont {Epstein}\ \emph {et~al.}(2016)\citenamefont
  {Epstein}, \citenamefont {Wong},\ and\ \citenamefont
  {Eleftheriades}}]{epstein2016cavity}%
  \BibitemOpen
  \bibfield  {author} {\bibinfo {author} {\bibfnamefont {A.}~\bibnamefont
  {Epstein}}, \bibinfo {author} {\bibfnamefont {J.~P.~S.}\ \bibnamefont
  {Wong}}, \ and\ \bibinfo {author} {\bibfnamefont {G.~V.}\ \bibnamefont
  {Eleftheriades}},\ }\bibfield  {title} {\enquote {\bibinfo {title}
  {Cavity-excited huygens’ metasurface antennas for near-unity aperture
  illumination efficiency from arbitrarily large apertures},}\ }\href@noop {}
  {\bibfield  {journal} {\bibinfo  {journal} {Nature communications}\ }\textbf
  {\bibinfo {volume} {7}},\ \bibinfo {pages} {10360} (\bibinfo {year}
  {2016})}\BibitemShut {NoStop}%
\bibitem [{\citenamefont {Asadchy}\ \emph {et~al.}(2016)\citenamefont
  {Asadchy}, \citenamefont {Albooyeh}, \citenamefont {Tcvetkova}, \citenamefont
  {D\'{\i}az-Rubio}, \citenamefont {Ra'di},\ and\ \citenamefont
  {Tretyakov}}]{Asadchy2016}%
  \BibitemOpen
  \bibfield  {author} {\bibinfo {author} {\bibfnamefont {V.~S.}\ \bibnamefont
  {Asadchy}}, \bibinfo {author} {\bibfnamefont {M.}~\bibnamefont {Albooyeh}},
  \bibinfo {author} {\bibfnamefont {S.~N.}\ \bibnamefont {Tcvetkova}}, \bibinfo
  {author} {\bibfnamefont {A.}~\bibnamefont {D\'{\i}az-Rubio}}, \bibinfo
  {author} {\bibfnamefont {Y.}~\bibnamefont {Ra'di}}, \ and\ \bibinfo {author}
  {\bibfnamefont {S.~A.}\ \bibnamefont {Tretyakov}},\ }\bibfield  {title}
  {\enquote {\bibinfo {title} {Perfect control of reflection and refraction
  using spatially dispersive metasurfaces},}\ }\href {\doibase
  10.1103/PhysRevB.94.075142} {\bibfield  {journal} {\bibinfo  {journal} {Phys.
  Rev. B}\ }\textbf {\bibinfo {volume} {94}},\ \bibinfo {pages} {075142}
  (\bibinfo {year} {2016})}\BibitemShut {NoStop}%
\bibitem [{\citenamefont {Epstein}\ and\ \citenamefont
  {Eleftheriades}(2016{\natexlab{a}})}]{Epstein2016_ieee}%
  \BibitemOpen
  \bibfield  {author} {\bibinfo {author} {\bibfnamefont {A.}~\bibnamefont
  {Epstein}}\ and\ \bibinfo {author} {\bibfnamefont {G.~V.}\ \bibnamefont
  {Eleftheriades}},\ }\bibfield  {title} {\enquote {\bibinfo {title} {Arbitrary
  power-conserving field transformations with passive lossless omega-type
  bianisotropic metasurfaces},}\ }\href {\doibase 10.1109/TAP.2016.2588495}
  {\bibfield  {journal} {\bibinfo  {journal} {IEEE Transactions on Antennas and
  Propagation}\ }\textbf {\bibinfo {volume} {64}},\ \bibinfo {pages}
  {3880--3895} (\bibinfo {year} {2016}{\natexlab{a}})}\BibitemShut {NoStop}%
\bibitem [{\citenamefont {Epstein}\ and\ \citenamefont
  {Eleftheriades}(2016{\natexlab{b}})}]{Epstein2016_prl}%
  \BibitemOpen
  \bibfield  {author} {\bibinfo {author} {\bibfnamefont {A.}~\bibnamefont
  {Epstein}}\ and\ \bibinfo {author} {\bibfnamefont {G.~V.}\ \bibnamefont
  {Eleftheriades}},\ }\bibfield  {title} {\enquote {\bibinfo {title} {Synthesis
  of passive lossless metasurfaces using auxiliary fields for reflectionless
  beam splitting and perfect reflection},}\ }\href {\doibase
  10.1103/PhysRevLett.117.256103} {\bibfield  {journal} {\bibinfo  {journal}
  {Phys. Rev. Lett.}\ }\textbf {\bibinfo {volume} {117}},\ \bibinfo {pages}
  {256103} (\bibinfo {year} {2016}{\natexlab{b}})}\BibitemShut {NoStop}%
\bibitem [{\citenamefont {Mohammadi~Estakhri}\ and\ \citenamefont
  {Al\`u}(2016)}]{Alu2016}%
  \BibitemOpen
  \bibfield  {author} {\bibinfo {author} {\bibfnamefont {N.}~\bibnamefont
  {Mohammadi~Estakhri}}\ and\ \bibinfo {author} {\bibfnamefont
  {A.}~\bibnamefont {Al\`u}},\ }\bibfield  {title} {\enquote {\bibinfo {title}
  {Wave-front transformation with gradient metasurfaces},}\ }\href {\doibase
  10.1103/PhysRevX.6.041008} {\bibfield  {journal} {\bibinfo  {journal} {Phys.
  Rev. X}\ }\textbf {\bibinfo {volume} {6}},\ \bibinfo {pages} {041008}
  (\bibinfo {year} {2016})}\BibitemShut {NoStop}%
\bibitem [{\citenamefont {{Kuester}}\ \emph {et~al.}(2003)\citenamefont
  {{Kuester}}, \citenamefont {{Mohamed}}, \citenamefont {{Piket-May}},\ and\
  \citenamefont {{Holloway}}}]{Holloway2003_GSTC}%
  \BibitemOpen
  \bibfield  {author} {\bibinfo {author} {\bibfnamefont {E.~F.}\ \bibnamefont
  {{Kuester}}}, \bibinfo {author} {\bibfnamefont {M.~A.}\ \bibnamefont
  {{Mohamed}}}, \bibinfo {author} {\bibfnamefont {M.}~\bibnamefont
  {{Piket-May}}}, \ and\ \bibinfo {author} {\bibfnamefont {C.~L.}\ \bibnamefont
  {{Holloway}}},\ }\bibfield  {title} {\enquote {\bibinfo {title} {Averaged
  transition conditions for electromagnetic fields at a metafilm},}\ }\href
  {\doibase 10.1109/TAP.2003.817560} {\bibfield  {journal} {\bibinfo  {journal}
  {IEEE Transactions on Antennas and Propagation}\ }\textbf {\bibinfo {volume}
  {51}},\ \bibinfo {pages} {2641--2651} (\bibinfo {year} {2003})}\BibitemShut
  {NoStop}%
\bibitem [{\citenamefont {Epstein}\ and\ \citenamefont
  {Eleftheriades}(2016{\natexlab{c}})}]{Epstein:16}%
  \BibitemOpen
  \bibfield  {author} {\bibinfo {author} {\bibfnamefont {A.}~\bibnamefont
  {Epstein}}\ and\ \bibinfo {author} {\bibfnamefont {G.~V.}\ \bibnamefont
  {Eleftheriades}},\ }\bibfield  {title} {\enquote {\bibinfo {title} {Huygens'
  metasurfaces via the equivalence principle: design and applications},}\
  }\href {\doibase 10.1364/JOSAB.33.000A31} {\bibfield  {journal} {\bibinfo
  {journal} {J. Opt. Soc. Am. B}\ }\textbf {\bibinfo {volume} {33}},\ \bibinfo
  {pages} {A31--A50} (\bibinfo {year} {2016}{\natexlab{c}})}\BibitemShut
  {NoStop}%
\bibitem [{\citenamefont {D{\'\i}az-Rubio}\ \emph {et~al.}(2017)\citenamefont
  {D{\'\i}az-Rubio}, \citenamefont {Asadchy}, \citenamefont {Elsakka},\ and\
  \citenamefont {Tretyakov}}]{Tretyakov2017_NLM}%
  \BibitemOpen
  \bibfield  {author} {\bibinfo {author} {\bibfnamefont {A.}~\bibnamefont
  {D{\'\i}az-Rubio}}, \bibinfo {author} {\bibfnamefont {V.~S.}\ \bibnamefont
  {Asadchy}}, \bibinfo {author} {\bibfnamefont {A.}~\bibnamefont {Elsakka}}, \
  and\ \bibinfo {author} {\bibfnamefont {S.~A.}\ \bibnamefont {Tretyakov}},\
  }\bibfield  {title} {\enquote {\bibinfo {title} {From the generalized
  reflection law to the realization of perfect anomalous reflectors},}\ }\href
  {\doibase 10.1126/sciadv.1602714} {\bibfield  {journal} {\bibinfo  {journal}
  {Science Advances}\ }\textbf {\bibinfo {volume} {3}} (\bibinfo {year}
  {2017}),\ 10.1126/sciadv.1602714}\BibitemShut {NoStop}%
\bibitem [{\citenamefont {{Kwon}}(2018)}]{Kwon2018_NLM}%
  \BibitemOpen
  \bibfield  {author} {\bibinfo {author} {\bibfnamefont {D.}~\bibnamefont
  {{Kwon}}},\ }\bibfield  {title} {\enquote {\bibinfo {title} {Lossless scalar
  metasurfaces for anomalous reflection based on efficient surface field
  optimization},}\ }\href {\doibase 10.1109/LAWP.2018.2836299} {\bibfield
  {journal} {\bibinfo  {journal} {IEEE Antennas and Wireless Propagation
  Letters}\ }\textbf {\bibinfo {volume} {17}},\ \bibinfo {pages} {1149--1152}
  (\bibinfo {year} {2018})}\BibitemShut {NoStop}%
\bibitem [{\citenamefont {Ra'di}\ \emph {et~al.}(2017)\citenamefont {Ra'di},
  \citenamefont {Sounas},\ and\ \citenamefont {Al\`u}}]{Alu2017_metagr}%
  \BibitemOpen
  \bibfield  {author} {\bibinfo {author} {\bibfnamefont {Y.}~\bibnamefont
  {Ra'di}}, \bibinfo {author} {\bibfnamefont {D.~L.}\ \bibnamefont {Sounas}}, \
  and\ \bibinfo {author} {\bibfnamefont {A.}~\bibnamefont {Al\`u}},\ }\bibfield
   {title} {\enquote {\bibinfo {title} {Metagratings: Beyond the limits of
  graded metasurfaces for wave front control},}\ }\href {\doibase
  10.1103/PhysRevLett.119.067404} {\bibfield  {journal} {\bibinfo  {journal}
  {Phys. Rev. Lett.}\ }\textbf {\bibinfo {volume} {119}},\ \bibinfo {pages}
  {067404} (\bibinfo {year} {2017})}\BibitemShut {NoStop}%
\bibitem [{\citenamefont {Epstein}\ and\ \citenamefont
  {Rabinovich}(2017)}]{Epstein2017_mtg}%
  \BibitemOpen
  \bibfield  {author} {\bibinfo {author} {\bibfnamefont {A.}~\bibnamefont
  {Epstein}}\ and\ \bibinfo {author} {\bibfnamefont {O.}~\bibnamefont
  {Rabinovich}},\ }\bibfield  {title} {\enquote {\bibinfo {title} {Unveiling
  the properties of metagratings via a detailed analytical model for synthesis
  and analysis},}\ }\href {\doibase 10.1103/PhysRevApplied.8.054037} {\bibfield
   {journal} {\bibinfo  {journal} {Phys. Rev. Applied}\ }\textbf {\bibinfo
  {volume} {8}},\ \bibinfo {pages} {054037} (\bibinfo {year}
  {2017})}\BibitemShut {NoStop}%
\bibitem [{\citenamefont {Popov}\ \emph
  {et~al.}(2019{\natexlab{a}})\citenamefont {Popov}, \citenamefont {Boust},\
  and\ \citenamefont {Burokur}}]{Popov2019}%
  \BibitemOpen
  \bibfield  {author} {\bibinfo {author} {\bibfnamefont {V.}~\bibnamefont
  {Popov}}, \bibinfo {author} {\bibfnamefont {F.}~\bibnamefont {Boust}}, \ and\
  \bibinfo {author} {\bibfnamefont {S.~N.}\ \bibnamefont {Burokur}},\
  }\bibfield  {title} {\enquote {\bibinfo {title} {Constructing the near field
  and far field with reactive metagratings: Study on the degrees of freedom},}\
  }\href {\doibase 10.1103/PhysRevApplied.11.024074} {\bibfield  {journal}
  {\bibinfo  {journal} {Phys. Rev. Applied}\ }\textbf {\bibinfo {volume}
  {11}},\ \bibinfo {pages} {024074} (\bibinfo {year}
  {2019}{\natexlab{a}})}\BibitemShut {NoStop}%
\bibitem [{\citenamefont {Wong}\ and\ \citenamefont
  {Eleftheriades}(2018)}]{Eleftheriades2018}%
  \BibitemOpen
  \bibfield  {author} {\bibinfo {author} {\bibfnamefont {A.~M.~H.}\
  \bibnamefont {Wong}}\ and\ \bibinfo {author} {\bibfnamefont {G.~V.}\
  \bibnamefont {Eleftheriades}},\ }\bibfield  {title} {\enquote {\bibinfo
  {title} {Perfect anomalous reflection with a bipartite huygens'
  metasurface},}\ }\href {\doibase 10.1103/PhysRevX.8.011036} {\bibfield
  {journal} {\bibinfo  {journal} {Phys. Rev. X}\ }\textbf {\bibinfo {volume}
  {8}},\ \bibinfo {pages} {011036} (\bibinfo {year} {2018})}\BibitemShut
  {NoStop}%
\bibitem [{\citenamefont {Popov}\ \emph {et~al.}(2018)\citenamefont {Popov},
  \citenamefont {Boust},\ and\ \citenamefont {Burokur}}]{Popov2018}%
  \BibitemOpen
  \bibfield  {author} {\bibinfo {author} {\bibfnamefont {V.}~\bibnamefont
  {Popov}}, \bibinfo {author} {\bibfnamefont {F.}~\bibnamefont {Boust}}, \ and\
  \bibinfo {author} {\bibfnamefont {S.~N.}\ \bibnamefont {Burokur}},\
  }\bibfield  {title} {\enquote {\bibinfo {title} {Controlling diffraction
  patterns with metagratings},}\ }\href {\doibase
  10.1103/PhysRevApplied.10.011002} {\bibfield  {journal} {\bibinfo  {journal}
  {Phys. Rev. Applied}\ }\textbf {\bibinfo {volume} {10}},\ \bibinfo {pages}
  {011002} (\bibinfo {year} {2018})}\BibitemShut {NoStop}%
\bibitem [{\citenamefont {Popov}\ \emph {et~al.}(2020)\citenamefont {Popov},
  \citenamefont {Burokur},\ and\ \citenamefont {Boust}}]{Popov2020}%
  \BibitemOpen
  \bibfield  {author} {\bibinfo {author} {\bibfnamefont {V.}~\bibnamefont
  {Popov}}, \bibinfo {author} {\bibfnamefont {S.~N.}\ \bibnamefont {Burokur}},
  \ and\ \bibinfo {author} {\bibfnamefont {F.}~\bibnamefont {Boust}},\
  }\bibfield  {title} {\enquote {\bibinfo {title} {Conformal sparse
  metasurfaces for wavefront manipulation},}\ }\href@noop {} {\  (\bibinfo
  {year} {2020})},\ \Eprint {http://arxiv.org/abs/2001.09878} {arXiv:2001.09878
  [physics.app-ph]} \BibitemShut {NoStop}%
\bibitem [{\citenamefont {{Kennedy}}\ and\ \citenamefont
  {{Eberhart}}(1995)}]{Kennedy_PSO}%
  \BibitemOpen
  \bibfield  {author} {\bibinfo {author} {\bibfnamefont {J.}~\bibnamefont
  {{Kennedy}}}\ and\ \bibinfo {author} {\bibfnamefont {R.}~\bibnamefont
  {{Eberhart}}},\ }\bibfield  {title} {\enquote {\bibinfo {title} {Particle
  swarm optimization},}\ }in\ \href {\doibase 10.1109/ICNN.1995.488968} {\emph
  {\bibinfo {booktitle} {Proceedings of ICNN'95 - International Conference on
  Neural Networks}}},\ Vol.~\bibinfo {volume} {4}\ (\bibinfo {year} {1995})\
  pp.\ \bibinfo {pages} {1942--1948 vol.4}\BibitemShut {NoStop}%
\bibitem [{\citenamefont {Oppenheim}\ and\ \citenamefont
  {Schafer}(1989)}]{Oppenheim1989_DFT}%
  \BibitemOpen
  \bibfield  {author} {\bibinfo {author} {\bibfnamefont {A.~V.}\ \bibnamefont
  {Oppenheim}}\ and\ \bibinfo {author} {\bibfnamefont {R.~W.}\ \bibnamefont
  {Schafer}},\ }\href@noop {} {\emph {\bibinfo {title} {Discrete-time Signal
  Processing}}}\ (\bibinfo  {publisher} {Prentice-Hall, Inc.},\ \bibinfo
  {address} {Upper Saddle River, NJ, USA},\ \bibinfo {year} {1989})\BibitemShut
  {NoStop}%
\bibitem [{\citenamefont {Popov}\ \emph
  {et~al.}(2019{\natexlab{b}})\citenamefont {Popov}, \citenamefont {Yakovleva},
  \citenamefont {Boust}, \citenamefont {Pelouard}, \citenamefont {Pardo},\ and\
  \citenamefont {Burokur}}]{Popov2019_LPA}%
  \BibitemOpen
  \bibfield  {author} {\bibinfo {author} {\bibfnamefont {V.}~\bibnamefont
  {Popov}}, \bibinfo {author} {\bibfnamefont {M.}~\bibnamefont {Yakovleva}},
  \bibinfo {author} {\bibfnamefont {F.}~\bibnamefont {Boust}}, \bibinfo
  {author} {\bibfnamefont {J-L.}\ \bibnamefont {Pelouard}}, \bibinfo {author}
  {\bibfnamefont {F.}~\bibnamefont {Pardo}}, \ and\ \bibinfo {author}
  {\bibfnamefont {S.~N.}\ \bibnamefont {Burokur}},\ }\bibfield  {title}
  {\enquote {\bibinfo {title} {Designing metagratings via local periodic
  approximation: From microwaves to infrared},}\ }\href {\doibase
  10.1103/PhysRevApplied.11.044054} {\bibfield  {journal} {\bibinfo  {journal}
  {Phys. Rev. Applied}\ }\textbf {\bibinfo {volume} {11}},\ \bibinfo {pages}
  {044054} (\bibinfo {year} {2019}{\natexlab{b}})}\BibitemShut {NoStop}%
\bibitem [{\citenamefont {Rabinovich}\ \emph {et~al.}(2019)\citenamefont
  {Rabinovich}, \citenamefont {Kaplon}, \citenamefont {Reis},\ and\
  \citenamefont {Epstein}}]{Epstein2019_mtg_exp}%
  \BibitemOpen
  \bibfield  {author} {\bibinfo {author} {\bibfnamefont {O.}~\bibnamefont
  {Rabinovich}}, \bibinfo {author} {\bibfnamefont {I.}~\bibnamefont {Kaplon}},
  \bibinfo {author} {\bibfnamefont {J.}~\bibnamefont {Reis}}, \ and\ \bibinfo
  {author} {\bibfnamefont {A.}~\bibnamefont {Epstein}},\ }\bibfield  {title}
  {\enquote {\bibinfo {title} {Experimental demonstration and in-depth
  investigation of analytically designed anomalous reflection metagratings},}\
  }\href {\doibase 10.1103/PhysRevB.99.125101} {\bibfield  {journal} {\bibinfo
  {journal} {Phys. Rev. B}\ }\textbf {\bibinfo {volume} {99}},\ \bibinfo
  {pages} {125101} (\bibinfo {year} {2019})}\BibitemShut {NoStop}%
\bibitem [{pro()}]{probe}%
  \BibitemOpen
  \href@noop {} {}\bibinfo {howpublished} {For further details on the
  EFS-105-12 fiber optic probe:
  \url{http://www.enprobe.de/products.htm}}\BibitemShut {NoStop}%
\bibitem [{\citenamefont {Abbe}(1873)}]{abbe1873beitrage}%
  \BibitemOpen
  \bibfield  {author} {\bibinfo {author} {\bibfnamefont {E.}~\bibnamefont
  {Abbe}},\ }\bibfield  {title} {\enquote {\bibinfo {title} {Beitr{\"a}ge zur
  theorie des mikroskops und der mikroskopischen wahrnehmung},}\ }\href@noop {}
  {\bibfield  {journal} {\bibinfo  {journal} {Archiv f{\"u}r mikroskopische
  Anatomie}\ }\textbf {\bibinfo {volume} {9}},\ \bibinfo {pages} {413--418}
  (\bibinfo {year} {1873})}\BibitemShut {NoStop}%
\bibitem [{\citenamefont {Novotny}\ and\ \citenamefont
  {Hecht}(2012)}]{novotny2012principles}%
  \BibitemOpen
  \bibfield  {author} {\bibinfo {author} {\bibfnamefont {L.}~\bibnamefont
  {Novotny}}\ and\ \bibinfo {author} {\bibfnamefont {B.}~\bibnamefont
  {Hecht}},\ }\href@noop {} {\emph {\bibinfo {title} {Principles of
  nano-optics}}}\ (\bibinfo  {publisher} {Cambridge university press},\
  \bibinfo {year} {2012})\BibitemShut {NoStop}%
\bibitem [{\citenamefont {Huang}\ \emph {et~al.}(2018)\citenamefont {Huang},
  \citenamefont {Qin}, \citenamefont {Liu}, \citenamefont {Ye}, \citenamefont
  {Qiu}, \citenamefont {Hong}, \citenamefont {Luk'yanchuk},\ and\ \citenamefont
  {Teng}}]{Qiu2018_diff_lens}%
  \BibitemOpen
  \bibfield  {author} {\bibinfo {author} {\bibfnamefont {K.}~\bibnamefont
  {Huang}}, \bibinfo {author} {\bibfnamefont {F.}~\bibnamefont {Qin}}, \bibinfo
  {author} {\bibfnamefont {H.}~\bibnamefont {Liu}}, \bibinfo {author}
  {\bibfnamefont {H.}~\bibnamefont {Ye}}, \bibinfo {author} {\bibfnamefont
  {C-W.}\ \bibnamefont {Qiu}}, \bibinfo {author} {\bibfnamefont
  {M.}~\bibnamefont {Hong}}, \bibinfo {author} {\bibfnamefont {B.}~\bibnamefont
  {Luk'yanchuk}}, \ and\ \bibinfo {author} {\bibfnamefont {J.}~\bibnamefont
  {Teng}},\ }\bibfield  {title} {\enquote {\bibinfo {title} {Planar diffractive
  lenses: Fundamentals, functionalities, and applications},}\ }\href {\doibase
  10.1002/adma.201704556} {\bibfield  {journal} {\bibinfo  {journal} {Advanced
  Materials}\ }\textbf {\bibinfo {volume} {30}},\ \bibinfo {pages} {1704556}
  (\bibinfo {year} {2018})}\BibitemShut {NoStop}%
\end{thebibliography}%

\end{document}